\documentclass[final,1p,times]{elsarticle}

\usepackage{graphicx}
\graphicspath{{images/}}
\usepackage{caption}
\usepackage{subcaption}
\usepackage{booktabs}
\usepackage{float}
\usepackage{amssymb}
\usepackage{amsmath}
\usepackage{bm}
\usepackage{color}
\usepackage{array}
\usepackage{algorithm}
\usepackage{algpseudocode}
\usepackage{hyperref}
\usepackage[capitalise]{cleveref}
\usepackage{mathtools}
\DeclarePairedDelimiterX{\norm}[1]{\lVert}{\rVert}{#1}

\DeclareMathOperator*{\argmin}{arg\,min}

\makeatletter
\renewcommand{\section}{\@startsection {section}{1}{\z@}%
  {-3.5ex \@plus -1ex \@minus -.2ex}%
  {2.3ex \@plus .2ex}%
  {\Large\scshape\bfseries}}

\renewcommand{\subsection}{\@startsection{subsection}{2}{\z@}%
  {-3.25ex\@plus -1ex \@minus -.2ex}%
  {1.5ex \@plus .2ex}%
  {\normalsize\bfseries}}
\renewcommand{\subsubsection}{\@startsection{subsubsection}{2}{\z@}%
  {-3.25ex\@plus -1ex \@minus -.2ex}%
  {1.5ex \@plus .2ex}%
  {\normalsize\bfseries}}           
\makeatother

\biboptions{comma,square,sort&compress}

\journal{Composites Structures}

\begin{document}

\begin{frontmatter}


  \title{A Bayesian Framework for Assessing the Strength Distribution of Composite Structures with Random Defects}

    \author[exe]{A. Sandhu}

  \author[tum]{A. Reinarz}

  \author[exe]{T. J. Dodwell \corref{cor1}}
  \ead{t.dodwell@exeter.ac.uk}

  \cortext[cor1]{Corresponding author}
  

  \address[exe]{College of Engineering, Mathematics  and Physical Sciences,  University  of Exeter, Exeter, EX4 4QF, UK.}

  \address[tum]{Institute of Informatics, Technical University of Munich, Boltzmannstr. 3, 85748 Garching, Germany.}

  \begin{abstract}
    
This paper presents a novel stochastic framework to quantify the knock down in strength from out-of-plane wrinkles at the coupon level. The key innovation is a Markov Chain Monte Carlo algorithm which rigorously derives the stochastic distribution of wrinkle defects directly informed from image data of defects. The approach significantly reduces uncertainty in the parameterization of stochastic numerical studies on the effects of defects. To demonstrate our methodology, we present an original stochastic study to determine the distribution of strength of corner bend samples with random out-plane wrinkle defects. The defects are parameterized by stochastic random fields defined using Karhunen-Lo\'{e}ve (KL) modes. The distribution of KL coefficients are inferred from misalignment data extracted from B-Scan data using a {\em modified} version of Multiple Field Image Analysis. The strength distribution is estimated, by embedding wrinkles into high fidelity FE simulations using the high performance toolbox  \texttt{dune-composites} from which we observe severe knockdowns of $74\%$ with a probability of $1/200$. Supported by the literature our results highlight the strong correlation between maximum misalignment and knockdown in coupon strength. This observations allows us to define a surrogate model providing fast assessment of predicted strength informed from stochastic simulations utilizing both observed wrinkle data and high fidelity finite element models.

  \end{abstract}


\end{frontmatter}


\section{Introduction}\label{sec:introduction}

While manufacturing large, complex composite components, small process-induced defects can form \cite{potter09}, for example porosity \cite{purslow1984}, in-plane fibre waviness \cite{liu2004compressive}, out-of-plane wrinkles \cite{dodders,mukhopadhyay}. In practice, we observe a distribution of locations, sizes and shapes of these defects, and therefore the direct effect they have on part performance is uncertain. Within composite aerospace industry, where safety is paramount, this uncertainty is mitigated by heuristic safety factors derived from extensive testing, which leads to high certification cost and over-conservativeness of design.  A modelling based initiative \cite{ic} allows numerical simulation and stochastic methods to be used in the certification process with the ultimate aim of lowering costs whilst challenging conservatism to obtain more optimized designs. In this work we develop a stochastic methodology to explore the distribution of strength of defective components by integrating finite element modelling of defects with observed measurement data about their size, location and morphology. Motivated by industry, we focus our study on out-of-plane defects, yet note that the general stochastic framework is applicable across a broad range of defects types, measurement data and modelling choices.

Wrinkle defects occur in the consolidation, forming and/or curing stages. There are number of different mechanisms that cause out-of-plane wrinkling \cite{lightfoot2013,dodwell,boisse2015}, although in most cases they are caused by the combination of the mechanics of the laminate in its uncured state, and the geometric constraints imposed by the manufacturing tool to which the laminate must conform. Importantly, the presence of a wrinkle defect can significantly effect the structural integrity of the as-manufactured part, in some cases leading to expensive wholesale rejection. Naturally there has been a focused research effort to develop non-destructive methods (NDT) to measure and classify wrinkles in as-manufactured parts \cite{sutcliffe2012,smith2013,nikishkov2013,meola2015,mizukami2016}. Amongst the variety of NDT methods available, only some are suitable for investigating geometric features at the meso-scale (sub-laminate scale). The most popular of these are X-ray computed tomography (XRCT) \cite{sutcliffe2012}, ultrasonic techniques \cite{smith2013,nikishkov2013} or infrared thermography \cite{meola2015}. XRCT can provide fibre scale resolution in a 3D volume but it is a much slower process compared to some ultrasonic methods which sacrifice accuracy for speed by limiting resolution. Faster NDT methods such as infrared thermography \cite{mizukami2016} enable scanning of larger areas but are limited to 2D information since they cannot penetrate much deeper than the surface of composite components. For these reasons, the industry prefers using ultrasonic techniques. Besides speed and accuracy, choice of NDT methods is driven by physical constraints. Aerospace components are usually too large to accommodate in a CT scanner. Ultrasonic methods require that parts be submerged in a coupling medium (typically water) which can again, be a limitation for large parts. Phased arrays \cite{meola2015} provide a viable alternative in such cases by enabling in-situ ultrasonic scanning. Principally, it is a combination of multiple individual ultrasonic probes programmed to work harmoniously, steering and focusing sound without source motion. The array of probes is embedded in flexible housing capable of bending along curved surfaces. The phased array functions like a medical ultrasound scanner with one key difference - different regions are explored by steering the ultrasonic beam instead of the device itself.

These wrinkle measurements have supported a growing research interest in the mechanical knockdown of wrinkle defects. The formation of wrinkles not only disrupts the even distribution of fibre and resin, but can significantly increase interply shear stresses triggering failure at significantly reduced loads \cite{lemanski2013,mukhopadhyay,fletcher2016,dunecomposites}. Numerical studies have used wrinkle measurements, to perform parametric studies using finite elements to explore the deterministic effect of variations of wrinkle shape on structural integrity \cite{mukhopadhyay,lemanski2013,xie2018}. These studies include prediction of both, failure initiation \cite{fletcher2016,dunecomposites} and mix-mode propagation \cite{mukhopadhyay}. Notably, in a recent study, Xie \emph{et. al.} explores the compressive strength of flat plate coupons containing internal wrinkles \cite{xie2018}. Here, the authors define wrinkles by a cosine function enveloped within a 3D Gaussian exponential. They use six classifying parameters to characterize a wrinkle namely, amplitude, wavelength, maximum misalignment angle, wrinkle-centre location, offset parameter for the cosine basis function and the extent of the Gaussian envelope. Based on a large number of simulations, the study recommends that  maximum wrinkle angle is the strongest indicator of strength knock down. Previously, Wang \emph{et. al.} developed three methods of fabricating out-of-plane waviness which were used to quantify a  wider class of wrinkles by introducing additional parameters \cite{wang2012}. However, those wrinkles are also constructed from cosine functions fitted to empirical misalignment measurements. Based on this study, Lemanski \emph{et. al.} illustrated numerically, a reduction of approximately $54\%$ for peak misalignment of $22^{\circ}$ \cite{lemanski2013}. Other studies have explored various combinations of these parameters, reporting compressive strength knockdown dependencies on other parameters such as amplitude \cite{elhajjar2014} and amplitude-wavelength ratio \cite{ohare1993}. 

In this contribution we identify and challenge two limitations in the existing numerical studies.  Firstly, the parameterization of the wrinkle has mostly been limited to single sinusoidal functions engulfed by a Gaussian envelope. We are aware of just one study by Kratmann \emph{et. al.} \cite{kratmann2009} which introduced a Fourier basis. This basis has limitations since a large number of Fourier modes are required to capture localized wrinkle profiles. We are unaware of any study that explores the sensitivity of their results to the choice of wrinkle parameterization. In this paper we seek a more general parameterization of wrinkles by exploiting the literature from Gaussian random fields \cite{spanos} and informing the parameterisation directly from measured data. Secondly, analytical studies have explored the effects of variations of out-of-plane wrinkles in a deterministic way. Current studies have not explored the stochastic effects of wrinkles, to derive a distribution in strength of components due to defects. The success of a stochastic simulation is dependent on the ability to define the probability distribution of possible wrinkles. In this paper we see this as a Bayesian question. What is the distribution of possible wrinkles given that we observe a set of wrinkles for which we have NDT measurements?

Bayesian statistics tools have been well developed in a broad range of application fields including groundwater hydrology \cite{marzouk2018,dodwell2015}, image visualization \cite{gilks1995} and ecology \cite{parno2014}. In this contribution we use these Bayesian tools to integrate high-fidelity finite element modelling capabilities of defective composites \cite{dunecomposites} with NDT measurements of wrinkles. The main idea is that from available information a broad probability distribution (the {\em prior} in the Bayesian terminology) is assigned to the input (wrinkle) parameters. If in addition we have measurement data related to real defects, it is possible to reduce the uncertainty and to get a better representation of defect parameterisation by conditioning the prior distribution on this data (leading to the {\em posterior}). However, direct sampling from a posterior distribution is not possible, therefore we generate samples using a Metropolis-Hastings type Markov chain Monte Carlo (MCMC) approach \cite{metropolis1953}. This approach consists of two main steps: (i) given the previous samples, a new sample is generated using a proposal distribution \cite{cotter2013}, such as a {\em random walk}; (ii) the likelihood of this new sample (i.e. how well the proposal matches observed defects) is compared to the likelihood of the previous sample. Based on this proposal and comparison steps, the proposed sample is either accepted and used for inference, or rejected and the previous sample is used again. The process leads to a Markov chain of possible defects, which have the probability distribution we seek, namely the distribution of wrinkles given observed measurement data. In our case these wrinkle samples can be embedded into a high-fidelity finite element model \cite{dunecomposites} to predict the strength of each sample in a Monte Carlo step. The output is the distribution of component strength given measurements of observed defects.

In this paper, we develop a framework to compute the statistics of the strength penalty imposed by defects in composites. We begin by introducing multi-disciplinary concepts that constitute this framework in \cref{sec:introduction}. The general description of framework is described in \cref{sec:bayesian} starting with wrinkle parameterization for a generic basis (\cref{sec:parameterization}), posterior sampling (\cref{sec:markovchain}) and finally, Monte Carlo simulations (\cref{sec:montecarlo}) to determine expected strength. An application of this framework is demonstrated through an industrial case study in \cref{sec:casestudy}, where we first describe the model problem (\cref{sec:modelproblem}), followed by the implementation of the method (\cref{sec:extraction,sec:definition,sec:femodelling}). Simulation results are presented in \cref{sec:results} alongside user inputs. In light of the findings and supporting literature \cite{xie2018,wang2012,lemanski2013} for slope-failure dependency, an engineer's model is proposed to predict failure based on maximum misalignment.

\section{Bayesian approach to construct  defect distributions from measured data}\label{sec:bayesian}

In this section we describe a Bayesian approach to construct a distribution of wrinkle defects from observed data, and how this distribution links to stochastic Monte Carlo simulations with a finite element model to predict the distribution of component strength. We have intentionally left the description general to show that the methodology works for a broad definition of wrinkle defects. In fact, provided an adequate basis is chosen, a variety of defects can be modelled within this framework. Here, a specific industrial motivated case study is considered in a later section.

\subsection{Parameterizing a wrinkle defect}\label{sec:parameterization}

\begin{figure}
  \centering
  \includegraphics[width = \linewidth]{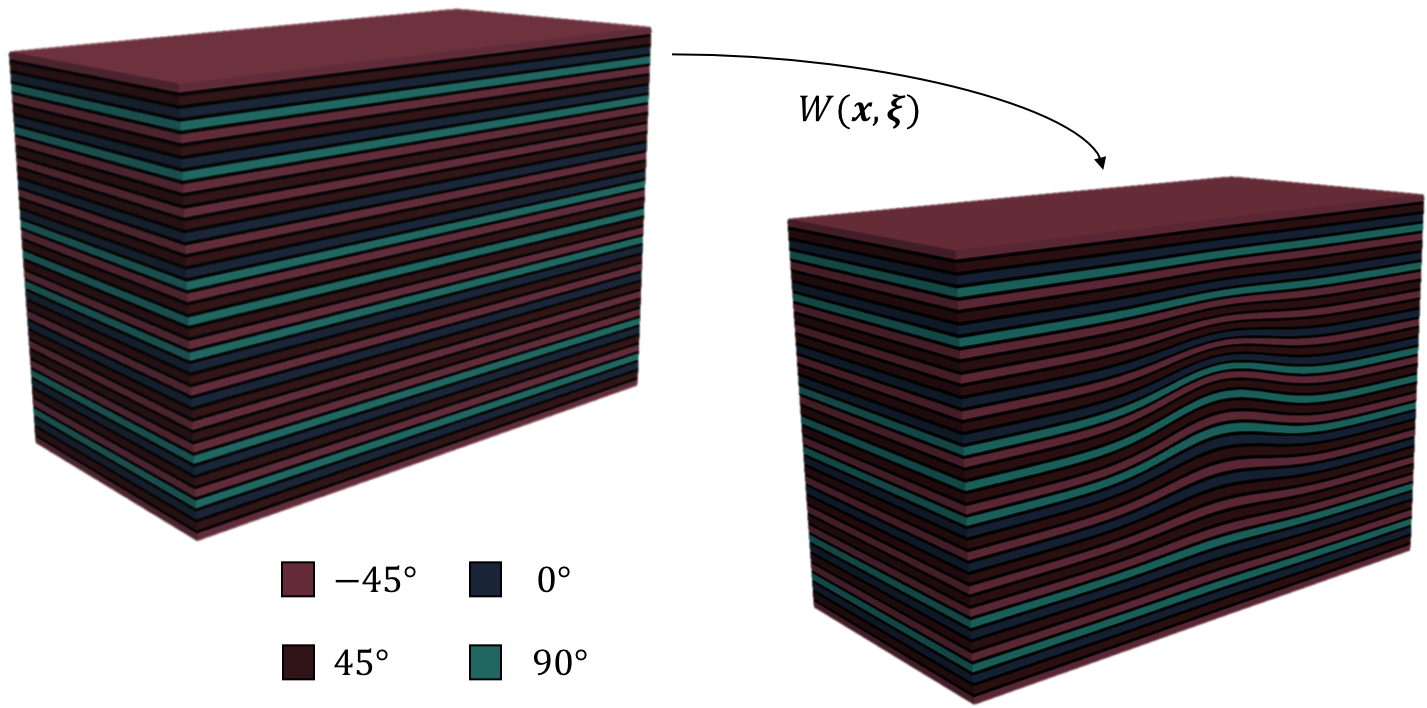}
  \caption{Illustration of \cref{eqn:wrinkleDef} showing the transformation from pristine to defective state for a 39 ply composite with a representative stacking sequence}
  \label{fig:wrinkleFig}
\end{figure}

An important step, and one that requires a modelling choice, is to define the method by which a wrinkle defect in a composite part is parameterized. In this contribution, and others, a wrinkle defect is defined by a deformation field $W:\Omega \rightarrow \mathbb R^3$ mapping a composite component from a pristine state (occupying $\Omega \subset \mathbb R^3$) to the defected state (occupying $W(\Omega)\subset \mathbb R^3$), Fig. \ref{fig:wrinkleFig}. To make our approach amenable to analysis we define this map by the finite dimensional representation
\begin{equation}\label{eqn:wrinkleDef}
  W(\mathbf{x}, \boldsymbol \xi) = \sum_{i=1}^{N}a_i \psi_i(\mathbf{x},{\bf b}), \quad \mbox{for} \quad \mathbf{x} \in \Omega \subset \mathbb R^3 \quad \mbox{and} \quad {\boldsymbol \xi} = [{\bf a},{\bf b}]^T \in \mathbb R^{N_w}.
\end{equation}
where the set $\{\psi_i({\bf x})\}$ define the orthonormal basis over which wrinkles are defined, and $\boldsymbol \xi$ is a vector of coefficients $N_w$ in length parameterizing the wrinkle. Here, we leave this choice open to show that the methodology presented is largely independent of parameterization of wrinkle since different choices have been made in the literature. Having said this, the choice of basis $\psi_i$ is important as it constrains the  representation of wrinkles. Therefore, it should be left as general as possible, with two important considerations
\begin{itemize}

\item The deformation induced by $W({\bf x}, \boldsymbol \xi)$ should not self-intersect. This is equivalent to the constraint that the $\det \mathcal J({\bf x},\boldsymbol \xi)) > 0$, for all ${\bf x} \in \Omega$, where $\mathcal J({\bf x},\boldsymbol \xi))$ is the Jacobian of the deformation map $W({\bf x},\boldsymbol \xi))$. At this stage it is sufficient to choose $\psi_i$ not self-intersecting,  and impose the constraint $\det \mathcal J({\bf x},\boldsymbol \xi) > 0$ during the posterior sampling (see below).

\item Since the data for which we tune our wrinkle distribution is estimated from approximations of the misalignment of plies, the basis functions $\psi_i({\bf x},{\bf b})$ should have well-defined first derivatives in the $x_1$ and $x_2$ directions. Moreover, the misalignment is computed as follows
  \begin{equation}
    \label{eq:misalignment}
    \tan \phi_j ({\bf x},\boldsymbol \xi) =  \sum_{i=1}^{N_w}a_i \frac{d\psi_i({\bf x},{\bf b})}{dx_j} \; \mbox{for} \quad j = 1 \; \mbox{and} \; 2.
  \end{equation}

\end{itemize}

\subsection{Posterior Sampling using a Metropolis-Hastings algorithm}\label{sec:markovchain}

Let the vector valued random variable $\boldsymbol \xi \in \mathbb X \subset \mathbb R^{N_w}$ denote the $N_w$-dimensional coefficient vector representing a random wrinkle profile. We will assume this has the general form defined by \eqref{eqn:wrinkleDef}. Let $\mathcal D_{obs} := \{ {\bf d}_{obs}^{(1)},\ldots,{\bf d}_{obs}^{(n)} \}$ denote the set of data measured from $n$ observed independent wrinkles, each characterized by ${\bf d}^{(i)}_{obs} = \{\phi_1^{(i)}, \phi_2^{(i)},\ldots,\phi_{N_\phi}^{(i)}\} \in \mathbb D \subset \mathbb R^{N_\phi}$. 

The {\em Forward Model} $F(\boldsymbol \xi) : \mathbb X \rightarrow \mathbb D$ maps a set of wrinkle coefficients $\boldsymbol \xi \in \mathbb X$ to the observable model output ${\bf d}_{obs}^{(i)} \in \mathbb D$. In this paper the observable data is the misalignment field of the wrinkle profile in the $x_1x_3$ plane, and therefore we define the Forward model as
\begin{equation}
  \label{eq:forwardmodel}
  \phi_j = \tan^{-1} \left( \sum_{i=1}^{N_w} \xi^{(i)} \frac{d \psi_i({\bf x_j})}{d x_1} \right) \quad \mbox{at measurement points} \quad {\bf x}_j \quad \mbox{for} \quad j = 1 \ldots N_\phi.
\end{equation}

In a Bayesian setting the first task is to construct the prior model and the likelihood function as probability distributions. The prior density is a stochastic model representing knowledge of the unknown $\boldsymbol \xi$ before Bayesian inversion on the data, denoted by the distribution $\pi_0(\boldsymbol \xi)$. The likelihood function specifies the probability density of the observation $\mathcal D_{obs}$ for a given set of parameters $\boldsymbol \xi$, denoted by $\mathcal L(\mathcal D_{obs}|\boldsymbol \xi)$. We assume that the data and the model parameters have the following stochastic relationship
\begin{equation}
  {\bf d}_{obs} = F(\boldsymbol \xi) + \boldsymbol \epsilon
\end{equation}
where the random vector $\boldsymbol \epsilon \in \mathbb R^{N_\phi}$ captures the measurement noise and other uncertainties in the observation-model relationship. Without additional knowledge of the measurement errors,   $\boldsymbol \epsilon$ is modelled as a zero mean Gaussian $\boldsymbol \epsilon \sim N(\boldsymbol 0,\Sigma_{\epsilon})$, for covariance $\Sigma_{\epsilon}$.

Let the misfit function for the observation ${\bf d}_{obs}^{(i)}$ be defined in the standard way, so that 
\begin{equation}
  \label{eq:misfit}
  \delta_i(\boldsymbol \xi) = \frac{1}{2} \norm[\Big]{ \Sigma_{\epsilon}^{-\frac{1}{2}} \left(F(\boldsymbol \xi) - \mathbf{d}_{obs}^{(i)}\right) }_2
\end{equation}
Whilst the misfit over the complete data set $\mathcal D_{obs}$ is defined by
\begin{equation}\label{eqn:newLike}
  \Delta(\boldsymbol \xi) = \min_{i = 1,\ldots,N_\phi} \delta_i(\boldsymbol \xi),
\end{equation}
which can be interpreted as mis-fit compared with the closest observed data point. The  likelihood function $\mathcal L(\mathcal D_{obs}|\boldsymbol \xi)$ is proportional to $\exp(-\Delta(\boldsymbol \xi))$, and by Bayes' formula, the posterior probability density is
\begin{equation}\label{eqn:post}
  \pi(\boldsymbol \xi|\mathcal D_{obs}) = \frac{1}{Z}\exp(-\Delta(\boldsymbol \xi)) \pi_0(\boldsymbol \xi),
\end{equation}
where $Z$ is a normalizing constant (which there is no need to compute).

The posterior distribution \eqref{eqn:post} can be sampled using MCMC methods such as the standard random walk algorithm \cite{cotter2013}. We now provide a brief review of the Metropolis-Hastings algorithm used in this contribution.
\begin{figure}
  \centering
  \includegraphics[width = \linewidth]{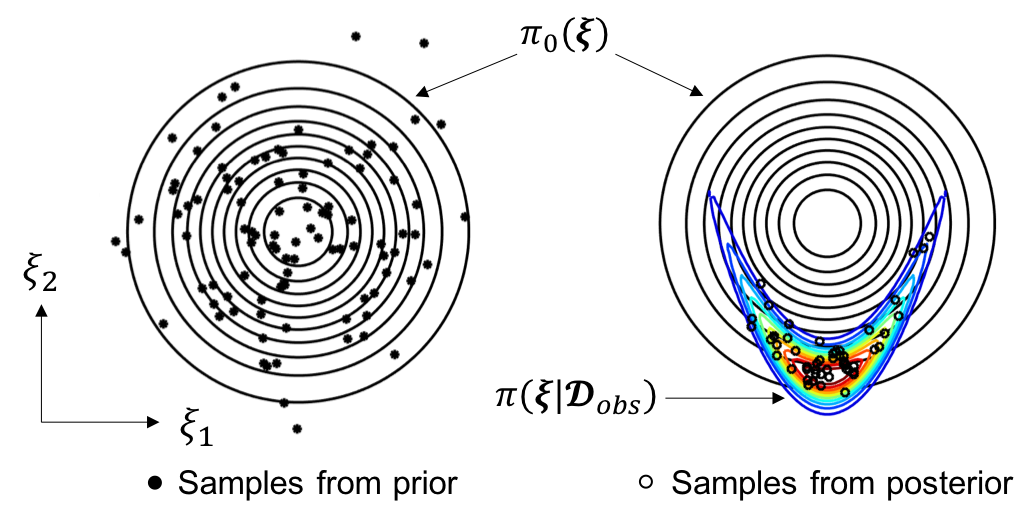}
  \caption{Representation of Bayesian approach in a simplified 2D parameter space for a known likelihood function (a variant of the Rosenbrock function). (Left) Sampling from isotropic Gaussian prior $\pi_0(\boldsymbol{\xi})$. (Right) MCMC sampling from posterior $\pi(\boldsymbol{\xi}|\mathcal{D}_{obs})$.}
  \label{fig:posterior_sampling}
\end{figure}
The first step is to define our prior distribution $\pi_0(\boldsymbol \xi)$ and pull a random sample as the starting point of our Markov Chain, say $\boldsymbol \xi^{(0)}$. Subsequent points on the Markov Chain $\boldsymbol{\xi}^{k}$ are generated by making a proposal $\boldsymbol{\xi'}$ defined by
\begin{equation}
  \label{eq:proposal}
  \boldsymbol{\xi'} = \sqrt{(1 - \beta^2)}\; \boldsymbol{\xi}^{k-1} + \beta \boldsymbol{\omega}
\end{equation}
This is a preconditioned Crank-Nicholson (PCN) proposal with $\beta \in \mathbb{R}$, a tuning parameter designed to enhance the efficiency of the standard Markov chain algorithm \cite{cotter2013}. The value $\beta$ controls the step size of a proposal. In \cref{eq:proposal}, $\boldsymbol{\omega} \in \mathbb{R}^{N_{w}}$ such that $\omega_j \sim \mathcal{N}(0,\sigma_{PCN}^2)$, are vectors of normally distributed random variables with standard deviation $\sigma_{PCN}$. The proposal $\boldsymbol{\xi'}$ is accepted for the next sample in the Markov chain $\boldsymbol{\xi}^{(k)}$ with the following probability
\begin{equation}
  \label{eq:acceptreject}
  \alpha(\boldsymbol{\xi'},\boldsymbol{\xi}^{k-1}) = \min \bigg\{ 1, \frac{\mathcal{L}(\mathcal D_{obs}|\boldsymbol{\xi'})\pi_0(\boldsymbol{\xi'})}{\mathcal{L}(\mathcal D_{obs}|\boldsymbol{\xi}^{k-1})\pi_0(\boldsymbol{\xi}^{k-1})} \bigg\} \quad \mbox{and} \quad \det \mathcal J({\bf x},\boldsymbol{\xi'})) > 0
\end{equation}
otherwise $\boldsymbol{\xi}^k = \boldsymbol{\xi}^{k-1}$. This process generates a series of samples which have the conditional probability distribution $\pi(\boldsymbol{\xi}|\mathcal D_{obs})$ which is the distribution of coefficients given the set of observations $\mathcal D_{obs}$.

Since we will use the $\boldsymbol{\xi}^{(k)}$ in Monte Carlo simulations we require $N_{MC}$ independent samples. Samples close to one another in a Markov chain are strongly correlated. By estimating the integrated autocorrelation time for each component of Markov Chain $\Lambda_i$ (see details in \cite[Ch. 5.8]{liu2008}) we can approximate a subsampling interval $\Lambda = \max (\Lambda_i)$ for which the samples are independent. Therefore in calculations which follow sampling only occurs after a \emph{burn-in} period of $b \gg \Lambda$ samples, to remove the influence of initial start of the chain $\boldsymbol{\xi}^{(0)}$ on the distribution of $\boldsymbol{\xi}$. Then samples are taken every $\Lambda$, generating the set of $N_{MC}$ independent random wrinkles from $\pi(\boldsymbol \xi|\mathcal D_{obs})$,
\begin{equation}
  \boldsymbol{\Xi} = \big\{\boldsymbol{\xi}^{(b + \Lambda)}, \boldsymbol{\xi}^{(b + 2\Lambda)}, \ldots, \boldsymbol{\xi}^{(b + N_{MC}\Lambda)} \big\}.\;\end{equation}

and $\Lambda$ is inversely related to $\beta$. However, the relationship between $\beta$ and acceptance ratio, another diagnostic property of MCMC, is more complex. Acceptance ratio is the ratio of the number of accepted proposals to the total number of a proposals made. Intuition suggests that a higher proposal density will lead to a higher percentage of wiser moves resulting in a greater acceptance ratio. A natural question then arises - what is the optimal proposal density or acceptance ratio? \citet{roberts1997} define a metric of efficiency (Langevin diffusion) in terms of acceptance ratio. This metric effectively quantifies the \emph{diffusion rate} of a chain through some unknown posterior distribution. Then the optimal acceptance ratio is one that maximizes the diffusion rate. The mathematical proof suggests that the asymptotically optimal acceptance ratio is approximately 0.25 however, in practice a ratio lower than 0.3 may be unachievable \cite{roberts1997}. The value of $\beta$ can be tuned to achieve this acceptance ratio.

The convergence of the chain to the posterior distribution \eqref{eqn:post}, can be monitored by running multiple independent parallel chains, and observing the convergence of $\mathbb E[\boldsymbol{\Xi}]$ (and perhaps higher moments) between all chains. Largely varying means of each chain would indicate the chains have yet to converge to a stationary distribution, and the burnin period should be extended.  This is particularly important in our case since with multiple independent observations $\mathcal D_{obs}$, the likelihood defined by \eqref{eqn:newLike} represents a multi-modal posterior with maxima at each data point. 

\subsection{Monte Carlo Simulations}
\label{sec:montecarlo}

Having generated a distribution of wrinkle profiles from observed data we are interested in computing the strength distribution of the defected components, by propagating these defects through a model and observing the distribution of component failure load. For this we introduce a finite element model $Q_M(\boldsymbol \xi):\mathbb X \rightarrow \mathbb R$ which maps a given wrinkle profile to an engineering quantity of interest, e.g. the expected load or moment at failure, or the probability of the failure occurring below a prescribed loading condition. The details of the particular finite element model and setup used in this contribution are provided in \cref{sec:femodelling}. The subscript $M$ indicates the number of degrees of freedom in that model, so that as $M \rightarrow \infty$ (under uniform mesh refinement) the expected value converges for some (inaccessible) random variable $Q:\mathbb X \rightarrow \mathbb R$, i.e. $\mathbb E[Q_M] \rightarrow \mathbb E[Q]$. We therefore seek to estimate
\begin{equation}
  \mathbb E[Q] = \int_{\mathbb X} Q(\boldsymbol \xi ) \pi(\boldsymbol \xi | \mathcal D_{obs})\;d\boldsymbol \xi.
\end{equation}
This can be estimated by $N_{MC}$ posterior samples $\boldsymbol \xi^{(i)} \sim\pi(\boldsymbol \xi | {\bf \mathcal D}_{obs})$ and the Monte Carlo estimate
\begin{equation}
  \hat Q_{M} = \frac{1}{N_{MC}}\sum_{i=1}^{N_{MC}} Q_M(\boldsymbol \xi^{(i)})
\end{equation}
which is a biased estimator with the mean square error
\begin{equation}
  \label{eq:meansquareerror}
  \varepsilon(\hat Q_{M})^2 = \mathbb E[Q - Q_M]^2  + \frac{\mathbb V(\hat Q_M)}{N_{MC}}, \quad \mbox{such that} \quad \mathbb V[\hat Q] \approx \frac{1}{N_{MC}-1}\sum_{i=1}^{N_{MC}} (Q_M(\boldsymbol \xi^{(i)}) - \hat Q_{M})^2.
\end{equation}

The first term in this expression represents the bias error in the model, arising from the quantity of interest $Q$ being approximated by a finite element calculation on a finite-dimensional grid. This error can be estimated from mesh analysis over a number of samples, as given in \cref{sec:femodelling}. The second term is the sampling error, arising from approximating $\mathbb E[Q_M]$ with only a finite number of samples. Care should be taken to balance these two errors to avoid unnecessary and expensive forward FE runs.

\section{Industrially Motivated Case study}\label{sec:casestudy}

  A case study based on industrial data is conducted to demonstrate the methodology described in \cref{sec:bayesian} and to introduce some bespoke but essential peripheral developments required to build theoretical models ($F(\boldsymbol \xi) : \mathbb{X} \rightarrow \mathbb{D}$ and $Q : \mathbb{X} \rightarrow \mathbb{R}$) from empirical data (B-scans). These developments are explained here in the context of a  model problem derived from an aircraft wing. The following section provides details and assumptions made in the model by briefly introducing the particular non-destructive testing (NDT) technique used to access internally contained wrinkles invisible from the outside. \cref{sec:extraction} then provides details on the Multiple Field Image Analysis (MFIA) algorithm that extracts alignment information from NDT results (B-scans). This is a necessary step prior to wrinkle parameterization as it constitutes the left hand side of \cref{eq:misalignment}. Next, we define an appropriate basis from which the right hand side of \cref{eq:misalignment} is derived to compute coefficients of best fit. Once parameterization is complete, all wrinkles considered here can be represented by some linear combination of the basis. Thus, in the posterior - the distribution of coefficients within the parameterized space $\mathbb{R}^{N_w}$ - one wrinkle only differs from another in the content of its coefficient vector, ${\bf a}$. It is important to understand that the method described in \cref{sec:bayesian} is directly applied within this parameterized space. Note, any mathematical operation applied to a \emph{wrinkle} hereafter, should be interpreted as a treatment of its corresponding coefficient vector.
  
  Now we wish to draw some conclusions about the posterior and model its evolution into a strength distribution in $\mathbb{R}$. An infinity of samples would be required to find the true distribution of wrinkles but we can produce useful results by pulling an appropriate number of independent and identically distributed samples or \emph{iid}s from the posterior. These iids or Monte Carlo samples are passed through an FE model $Q : \mathbb{X} \rightarrow \mathbb{R}$ that outputs a scalar strength. Details of the FE simulation are provided in \cref{sec:femodelling}. Continuous distributions can be approximated from the iids within some confidence bounds. We attempt to interpret the results thus generated in terms of Weibull statistics to understand the cumulative effects of misalignments. However, an engineering result of crucial importance is presented in light of the findings up to this point. It is a parameterized exponential relationship between the first derivative of a wrinkle, $W^\prime(\mathbf{x},\boldsymbol{\xi})$ and its failure moment. It offers a major time advantage by replacing lengthy FE calculations with an analytical formula.

\subsection{Model Problem and it's industrial application}\label{sec:modelproblem}

\begin{figure}[H]
  \centering
  \includegraphics[width=0.65\linewidth]{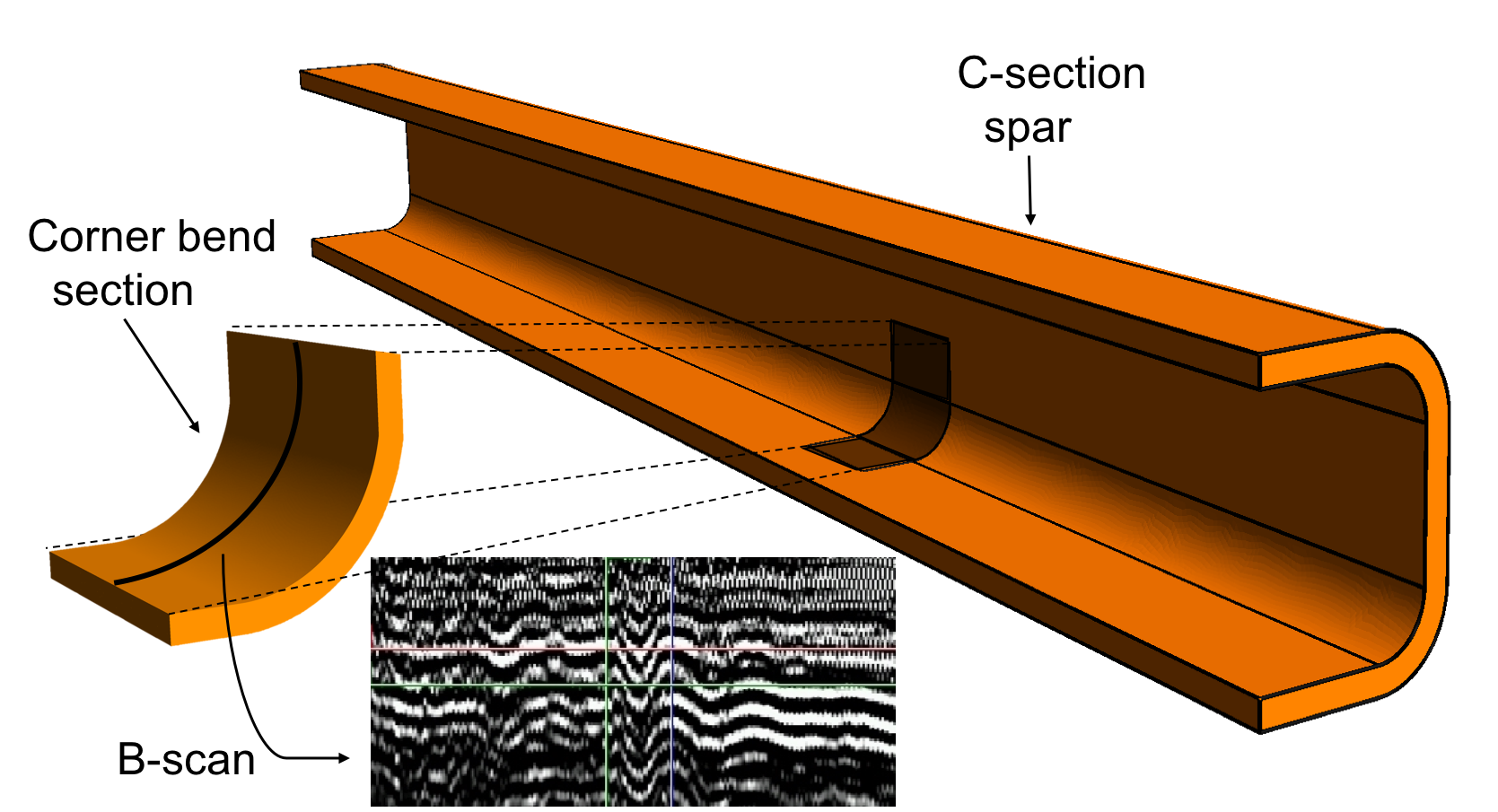}
  \includegraphics[width=0.34\linewidth]{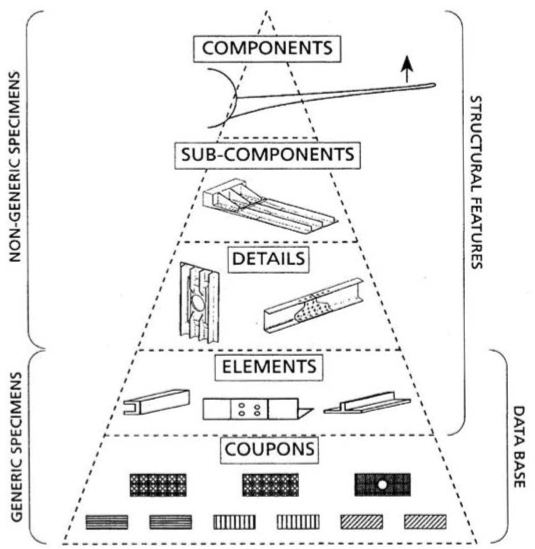}
  \caption{Schematic of a wing spar highlighting the region of showing a B-scan at a defect location.}
  \label{fig:cornerbend}
\end{figure}

\Cref{fig:cornerbend} (right) illustrates a typical aircraft development program followed by industry. As more exotic materials and/or technology are introduced into the aerospace industry, development costs continually increase. In the interest of preserving the health of the aviation economy, a new initiative encouraging modelling alongside physical testing is gaining traction. To achieve similar levels of robustness, the modelling track follows a conceptually similar development pyramid whereby the number of simulations conducted at the coupon level are much greater than components or systems. Motivated by this philosophy, one corner of a C-section composite (CFRP) wing spar is considered for the model problem. \Cref{fig:cornerbend} shows the corner bend coupon as a building block of a element-level part (spar) to clarify that this study explores the coupon level exclusively. Additional detail about the model problem is provided in \cref{fig:femodel}.

Upper and lower wing covers, bound together by the fore and aft spars form the fuel tanks in an aircraft. For a multitude of reasons, fuel must be stored under pressure which exerts an opening moment along the inner radius of the spar. During manufacture these regions (highlighted in \cref{fig:cornerbend}) are susceptible to wrinkle formation rendering them of critical importance for failure initiation. This case study focuses on characterization of such defects to simulate their effects on part strength using a corner bend sample representative of the region shown in \cref{fig:cornerbend}.

Visualizing and measuring the wrinkles that we intend to model can be particularly challenging since their parent components may be inconveniently large ($ > 10$m). To overcome problems like immersion in ultrasonic imaging, a \emph{phased array} is used for scanning. There are two major advantages of a phased array: 1) it eliminates motion of probes by using multiple sensors arranged so that they can be fired individually to allow beam steering and wavefront manipulation (focus) \cite{meola2015,phasedarray} and, 2) scanning can be conducted in-situ. Of greater importance perhaps, are the two key limitations since, in this case, they simplify our problem.

Firstly, the resultant image produced is a slice through thickness rather than a volume (see \cref{fig:cornerbend} (left)) meaning that we only have two dimensional information about a wrinkle. Due to the geometry of the spar, wrinkles form in a way that its span in $x_2$ is orders of magnitude greater than perturbations in $x_1$ and $x_3$. Therefore, within the vicinity of the B-scan, we can safely assume the wrinkle to be prismatic in $x_2$.

Secondly, the ultrasonic beams are focussed at a particular depth such that scans contain a corresponding high resolution region. Ply boundaries begin to fade into the surroundings further from the focussed band (see lower half of B-scan in \cref{fig:cornerbend}). Therefore, at a sampling location outside the focussed region, the existence of a global minimum in gray scale variance is not guaranteed. Since the wrinkle shows no clear signs of decay in $x_3$ within the subregion, we discard the $x_3$ coordinate of the sampling points. We have thus reduced the number of dependent variables of the alignment map to one.

\subsection{Extracting wrinkle data from B-Scans using Multiple Field Image Analysis (MFIA)}\label{sec:extraction}

A variety of image processing tools for investigating alignment exist, see for example the review by Smith \emph{et. al.} \cite{ndtreview} and other contributions \cite{phasedarray,smith2009}. In this contribution we use Multiple Field Image Analysis (MFIA) algorithm introduced by Creighton \emph{et. al.} \cite{Cre01} to estimate the misalignment of a wrinkled ply at a given position in the B-Scan image. We briefly review the method and describe some adaptations made to handle low resolution B-Scan images and the computational efficiency of the original approach \cite{Cre01}.

Multiple Field Image Analysis (MFIA) \cite{Cre01} uses a pixelated gray-scale image. At a given point ${\bf x} = (x_1,x_3)$, a {\em trial fibre} is introduced. This is an array of pixels of length $H$, centered about ${\bf x}$, and orientated at an angle $\theta$ to the $x_1$ or $x_2$ axis. At each sampling point, the algorithm finds the orientation $\theta$ of the trial fibre which minimizes variance in gray scale along it's length, i.e. the misalignment at point is the defined by
\begin{equation}
  \phi = \argmin\left(\mathcal J(\theta) \right), \quad \mbox{where} \quad \mathcal{J}(\theta) := \frac{1}{H}\int_{-H/2}^{H/2} ( \mathcal G(h,\theta) - \overline{\mathcal G}(\theta))^2 dh,
  \label{eq:variance}
\end{equation}
$\overline{\mathcal{G}}(\theta) = \frac{1}{H}\int_{-H/2}^{H/2}\mathcal G(h,\theta) dh$ is the mean gray-scale along the fibre, with $\mathcal G(h,\theta)$ defining the gray-scale at a point $(x_1 + h\cos\theta,x_3 + h\sin\theta)$  along its length. The procedure is repeated for an array of sample points $\mathbf{x}^{(k)} = (x_1^{(k)},x_3^{(k)})$ for $k \in \{1,2,\ldots, N_\phi\}$ where $N_\phi$ is number of pixels sampled per image.

\begin{figure}
  \centering
  \includegraphics[width=\linewidth]{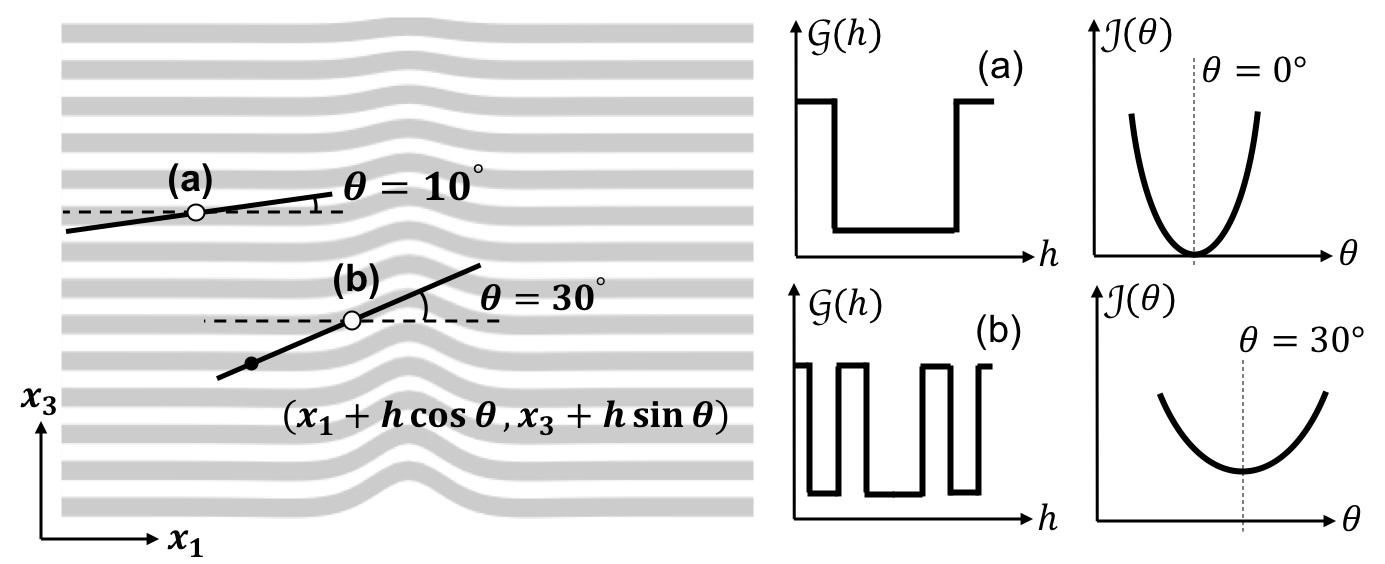}
  \caption{Estimating alignment at a point by minimizing the integral of the gray scale over the trial fibre using the MFIA algorithm \cite{Cre01}. Randomly sampled points are used to reconstruct an alignment over the domain.}
  \label{fig:mfia}
\end{figure}

A single point represents an individual optimization problem defined in \cref{eq:variance}. Sampling every pixel is prohibitively expensive, therefore we want a method to selectively distribute sampling points in such a way that concentrate evaluations to regions of misalignment. To do this we develop a hierarchical approach. The B-scan image is divided into a coarse rectangular mesh (level $j=0$) with $m_0$ cells. We generate a sequence of levels by uniformly refining the mesh, given $m_j = 4^jm_0$ on level $j$. The method starts by computing the misalignment $\phi({\bf x}_k^{(j)})$ at $n^{(i)}_j = \lceil N^{(j)}_\phi/m_j\rceil$ randomly sampled points in each cell, ${\bf x}_k^{(j)} \in \Omega_i$ for $k = 1, \ldots n^{(i)}_j$, $i = 1, \ldots, m_j$ and  $j=0$. In each cell $\Omega^{(i)}$ we compute the mean absolute misalignment
\begin{equation}
  \overline{\phi}^{(j)}_i = \frac{1}{|\mathcal X^j_i|}\sum_{{\bf x} \in \mathcal X^j_i} |\phi({\bf x})|, \quad \mbox{where the set is defined} \quad \mathcal X^{j}_i = \{{\bf x}_k : {\bf x}_k \in \Omega^{j}_i\}
\end{equation}
For each cell on level $j$ we compute the misalignment and its percentage contribution to the total mean absolute misalignment i.e.
\begin{equation}
  \gamma_i^{(j)} = \overline{\phi}^{(j)}_i / \sum_{c=1}^{m_j} \overline{\phi}^{(j)}_c
\end{equation}
For the next level $j+1$, $N_\phi^{(j+1)}$ more samples are taken for the four cells $\Omega^{(j+1)}$ created from subdividing $\Omega_i^{(j)}$ we take
\begin{equation}
  n_i^{(j+1)} = \lceil \gamma_i^{(j)} N_\phi^{(j)} \rceil 
\end{equation}
more samples where $\overline{\phi}_i$ is the average misalignment of the i\emph{th} cell. This is one of the key peripherals developed for the MFIA framework that allows us to generate a continuous alignment field from a manageable number of samples.

\begin{figure}
  \centering
  \includegraphics[width = 0.32\linewidth]{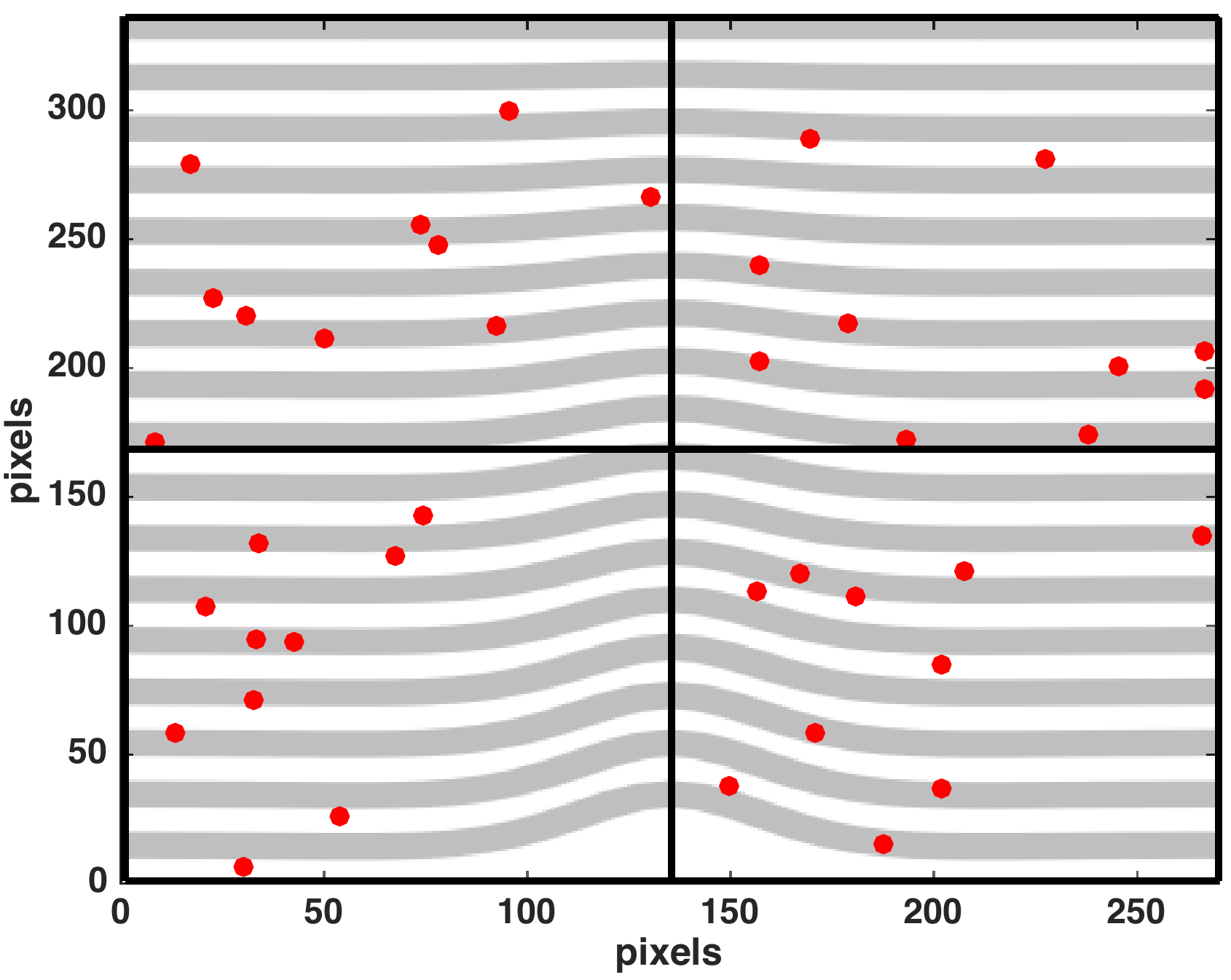}
  \includegraphics[width = 0.32\linewidth]{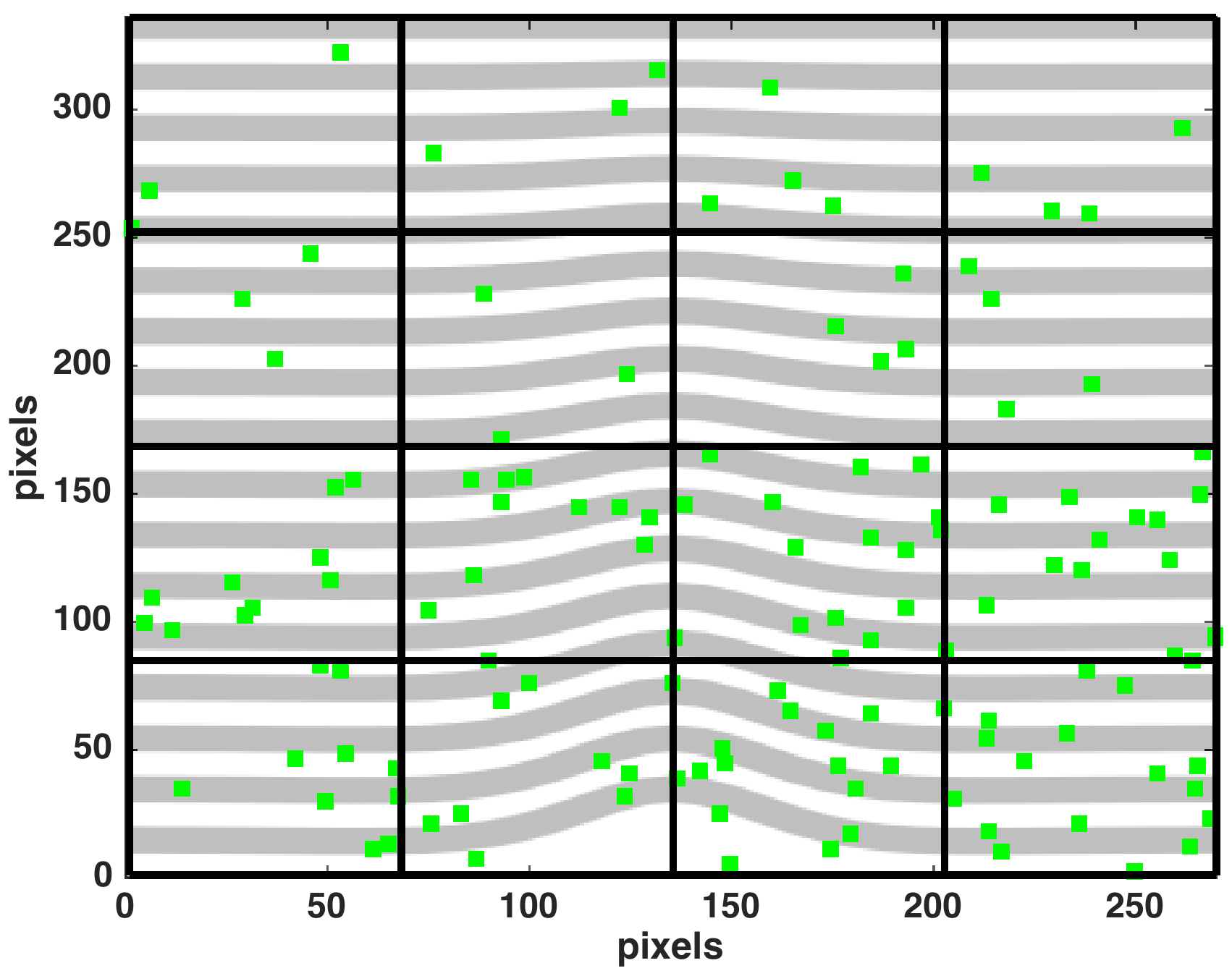}
  \includegraphics[width = 0.32\linewidth]{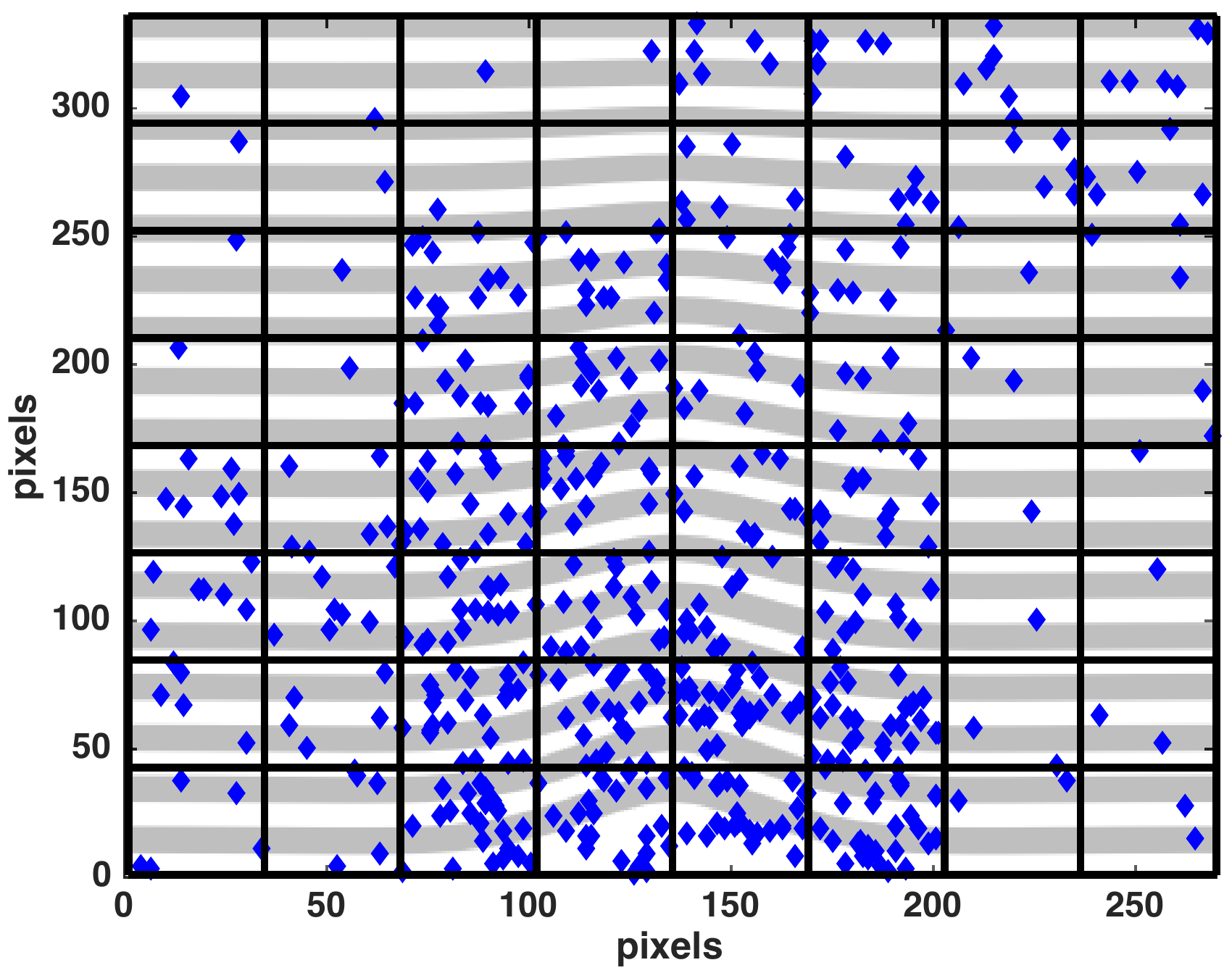}
  \caption{A hierarchical multilevel sampling scheme that is biased towards regions of high misalignment is used draw sampling locations for the trial fibre. The number of new samples per cell on every level are proportional to the relative average misalignment observed in that cell on the previous level. The figure shows level 1 through 3 from left to right and each picture only shows the samples drawn on that level.}
  \label{fig:sampling}
\end{figure}

\emph{Remark:} We note that MFIA is sensitive to trial fibre length $H$. In our experience, $H = 3t$ produces reliable results where $t$ is the average ply thickness in pixels. Moreover, these B-scans are flat representations of cornerbends thus requiring a geometric transformation to undo the apparent corner opening.

\subsection{Defining a Wrinkle, Prior and likelihood definition}\label{sec:definition}

In the methodology (\cref{sec:bayesian}), the parameterization of a wrinkle defect is left very general \eqref{eqn:wrinkleDef}. In this section we refine this definition towards the particular application and available data that we consider. The wrinkles are defined by the wrinkle functions
\begin{equation}
  \label{eq:stochastic}
  W({\bf x},\boldsymbol{\xi}) =  g_1(x_1)g_3(x_3)\sum_{i=1}^{N_w} a_i f_i(x_1,\lambda).
\end{equation}
where $g_i(x_i)$ are decay functions (defined in \cref{eqn:decayfunctions}), $f_i(x_1,\lambda)$ are the first $N_w$ Karhunen-Lo\'{e}ve (KL) modes parameterized by the length scale $\lambda$ and $a_i$ the amplitudes. In the results which follow both the amplitude modes and the length scale are taken as random variables, so that the stochastic vector is defined by $\boldsymbol \xi = [a_1,a_2,\ldots,a_{N_w},\lambda]^T$. We now briefly discussed the assumptions under which this choice of wrinkle function has been made.
\begin{itemize}

\item {\bf Prismatic in $x_2$}. The wrinkle function \cref{eq:stochastic} is assumed to have no $x_2$ dependency, and therefore the wrinkles are prismatic along the width. Inline with the ASTM standardization \cite{astm}, four point bend tests were conducted on $52$mm wide corner bend samples. In all four samples considered, wrinkles were prismatic over this coupon width.

\item {\bf  Karhunen-Lo\'{e}ve (KL) modes for wrinkle}. KL modes are widely use in generating random fields, since they generate random functions which display an underlying spatial correlation structure. We note other choices could also have been used e.g. piecewise cubic splines, wavelets or Fourier modes. The set of function $f_i(x_1,\lambda)$ are proportional the first $N_w$ (normalized) 1D eigenfunctions associated with the $N_w$ largest eigenvalues of the one-dimensional two-point, squared exponential covariance operator
  \begin{equation}
    C(x,y) = \sigma_f^2 \exp\left(-\frac{(x-y)^2}{\lambda^2}\right),\quad \mbox{for any} \quad  x, y \in \mathbb R
    \label{eq:covariance}
  \end{equation} 
  The normalizing constant for each mode is the square root of it's associated eigenvalues. Further information of KL modes and their use as random fields is widely available, see for example \cite{spanos}. We note that due to their natural ordering (in decreasing eigenvalue and wavelength) the modes can efficiently represent functions which display behaviour with a characteristic length scale $\lambda$ yet offer more flexibility than a simple choice such as $\sin(\lambda x_1)$. The runs which follow take $N_w = 30$, which was chosen since with an approximate value of $\lambda = 12.9$mm, higher KL modes give undulations on a wavelength shorter than pixels of the B-Scan. Furthermore, from the data we estimate $\sigma_f = 0.1425$, but note the output of the model is insensitive to the choice of this value.

\item {\bf Wrinkle Decay out side of B-Scan.} For the limited data we have, each B-Scan is centered to the midpoint of the corner radius $x_1^\star = R\pi/4$, and focuses at a fixed depth $x_3^\star = 4.8$mm. For the type of wrinkles considered here, no perturbations are visible on the inside or outside face of the component and the wrinkles were always localized to the corner radius. Again, without further data on their spatial statistics of the wrinkle distribution, we make the simplifying assumption of introducing decay function in both the $x_1$ and $x_3$ direction (as also considered in other publications \cite{xie2018,dunecomposites}, defined by
  \begin{equation}\label{eqn:decayfunctions}
    g_i(x_i) = \exp\left(-\left(\frac{x_i - x_i^\star}{\eta_{g_i}}\right)^n \right)
  \end{equation}
  For the simulations which follow we take $\eta_i = -(x_i^\star)^4/\log(10^{-6})$. This choice of $\eta_i$ gives the assumption that the wrinkle height is at most $10^{-6}$mm outside of the corner radius and on the inner/outer face. For both $x_1$ and $x_3$ directions, we choose the hyper-parameter $n = 4$. This value is selected to provide the best fit to the observed wrinkle profiles. Better decay functions can be derived from higher quality scans, however, little difference in output was observed for even values of $n>4$.
\end{itemize}

With more available data, which includes a broader class of wrinkle defects\cite{lightfoot2013,dodders,mukhopadhyay,dunecomposites}, this definition could be generalized. Yet, here the choice is sufficient to demonstrate the methodology, and draw some interesting preliminary engineering results.

Now that we have defined our parameterization of a wrinkle, it remains to define the prior distribution for the random parameters and the parameter $\Sigma_\epsilon$ for the misfit function. First, denoted $\pi_0(\boldsymbol \xi)$, we define the prior. This is done by analyzing each of four B-Scans and fitting the wrinkle function \eqref{eq:stochastic} in the least-squared sense using an optimizer (e.g. \texttt{fminsearch} in \textsc{Matlab} \cite{matlab}). This then provides just four values for each parameter. We approximate the prior as an independent multidimensional Gaussian distribution with mean taken over all measurements and a variance of all $4$ samples multiplied by the student t-test factor to account for uncertainty due to only four data points. We note that with just $4$ samples and a two sided confidence bounds of $95\%$ this is a factor of $3.18$. Secondly, in the definition of the misfit function \cref{eq:misfit}, we require the user-defined correlation matrix $\Sigma_\epsilon$, which defines the uncertainty in the measured data. In our case measurement error comes from two sources (1) the accuracy of B-Scan data and analysis method (MFIA) (2) the sampling error of the data since (in our case) we only have $4$ samples. 
To account for the first of these sources of measurement error we assume that all data points are accurate up to $\pm 2.5^\circ \approx \pm 0.044$rad. This was estimated from comparing MFIA outputs to micrographs of wrinkle sections. Further details are not given on how this is constructed, as it is well documented that the MCMC outputs are not sensitive to the fine scale accuracy of $\Sigma_\epsilon$. Secondly to account for sample data we rescale $\Sigma_\epsilon$ by the  student t-test factor to ($95\%$) confidence, which we denote $\tau_{N_\phi}$. We remark that $\tau_{N_\phi} \rightarrow 1$ and $N_\phi \rightarrow \infty$, and in this case $\Sigma_\epsilon$ is purely driven by the accuracy of the B-Scan data. For our example we therefore set $\Sigma_e = \tau_{N_\phi} 0.044\mathbb I$.

\subsection{Finite Element Modelling}
\label{sec:femodelling}

For each wrinkle sample generated using the MCMC approach, a finite element analysis is used to predict the corner bend strength (CBS) of that defected component. Finite element modelling was conducted using high performance finite element code \texttt{dune-composites} \cite{dunecomposites}. In the model, the curved laminates were assumed to have the nominal width of 52 mm. The plies were assumed to have a thickness of 0.24 mm, with a 0.015 mm interface layer of pure resin between each ply. This is based upon measurements taken from micrograph images of the curved laminates as described by Fletcher {\em et al.} \cite{fletcher2016}. The assumed mechanical properties for both the fibrous ply material and the resin rich interface material are given in \cref{tab:properties}. A discussion on how they have been chosen from various sources is given by Fletcher et al. \cite{fletcher2016}.

\begin{figure}[H]
  \centering
  \includegraphics[width=\linewidth]{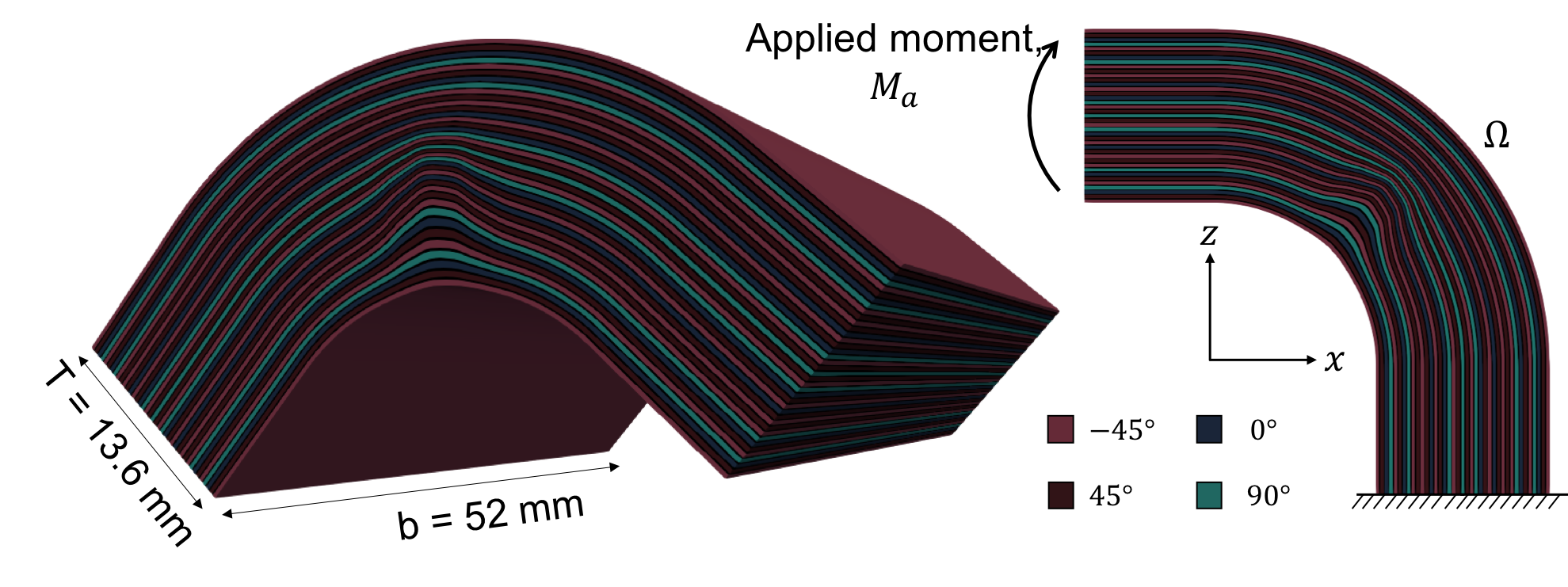}
  \caption{FE model showing the true geometry of the part with a sample wrinkle amplified for visual clarity. Note that it is a fully internal wrinkle with no trace at the surfaces.}
  \label{fig:femodel}
\end{figure}

Modelling the full 3D bending test (according the ASTM standard \cite{astm}) with rollers and contact analysis would be extremely computationally expensive. Therefore a simplified model was used. Curved laminates were modelled with shortened limbs; of length $10$mm, approximately equal to the thickness of the laminate. A unit moment was applied to the end of one limb using a multi-point constraint (MPC), with all degrees of freedom fixed at the end of the opposite limb. Whilst this does not accurately model stresses in the limbs, it gives the same stress field towards the apex of the curved section as a full model with rollers. In this region there is a pure moment (without shear) caused by the roller displacement. Since this is the critical region where both wrinkles and failure occurs during the tests, it implies the simplified model is suitable for predicting CBS. The setup of the model is summarized in \cref{fig:femodel}.

Each finite element model contains approximately 1.1 million 3D 20-node serendipity elements, with 8 elements per ply thickness and 4 in the interply regions adding up to roughly 2 million nodes (or $6$ million degrees of freedom). This model resolution follows from the mesh convergence study as presented by Reinarz {\em et al.} \cite{dunecomposites}. Failure of the coupon is measured according to Camanho's failure criterion \cite{Camanho}, whereby a numeric value is assigned to a particular combination of peak tensile and shear stresses. 
\begin{equation}
  \label{eq:failure}
  \mathcal F(\sigma) = \sqrt{\bigg(\frac{\sigma_{33}^+}{s_{33}}\bigg)^2 + \bigg(\frac{\sigma_{23}}{s_{23}}\bigg)^2 + \bigg(\frac{\sigma_{13}}{s_{13}}\bigg)^2}
\end{equation}
Here the subscripts denote the direction of stresses in local coordinates and $s_{ij}$ denotes allowable stresses. Note that $\sigma_{33}^+$ is set to 0 if the stress component is negative. Failure occurs when $\mathcal F(\sigma) = 1$. Usually, failure of a system such as \cref{fig:femodel} occurs due to delamination which indicates that the peak stresses is likely occur in the resin rich interply regions. Fletcher \emph{et. al.} have shown that the Camanho failure criterion predicts failure to within $5\%$ of average experimental test values \cite{fletcher2016} with treated edges to mitigate premature failure. Here, for simplicity, we discount the edge effects by not evaluating the failure criterion close to the boundary to isolate the effects of wrinkles. More precisely, $\mathcal{F}(\sigma)$ is evaluated within a subregion such that $x_2 \in [15$mm$,37$mm$]$ which is $15$mm away from each edge.

\begin{table}
  \centering
  \label{tab:properties}
  \begin{tabular}{ll|llll}
    \toprule
    \multicolumn{2}{c}{\textbf{Geometry}} & \multicolumn{2}{c}{\textbf{Ply properties}} & \multicolumn{2}{c}{\textbf{Resin properties}} \\ \cmidrule(r){1-2} \cmidrule(l){3-6}
    number of plies                  & 39         & $E_{11}$                  & 162 GPa         & $E$                        & 10 GPa           \\
    radius                     & 22 mm         & $E_{22}, E_{33}$          & 10 GPa          & $\nu$                      & 0.35             \\ 
    limb length                & 10 mm         & $G_{12}, G_{13}$          & 5.2 GPa         & \multicolumn{2}{c}{\textbf{Allowables}}       \\ \cmidrule(l){5-6}
    ply thickness              & 0.24 mm         & $G_{23}$                  & 3.5 GPa         & $s_{13}, s_{23}$           & 97 MPa           \\
    interply thickness         & 0.015 mm      & $\nu$                     & 0.35            & $s_{33}$                   & 61 MPa           \\ \bottomrule
  \end{tabular}
  \caption{Assumed mechanical properties for CFRP material (M21/IMA), where 1 is the fibre direction in-plane, 2 is perpendicular to the fibre direction in-plane and 3 is out-of-plane. $s_{33}$ is the tensile through-thickness strength and $s_{13}$ is the transverse shear strength.}
\end{table}

\section{Results}\label{sec:results}

\subsection{Bayesian Sampling of wrinkles}

\begin{figure}
  \centering
  \includegraphics[width = 0.6\linewidth]{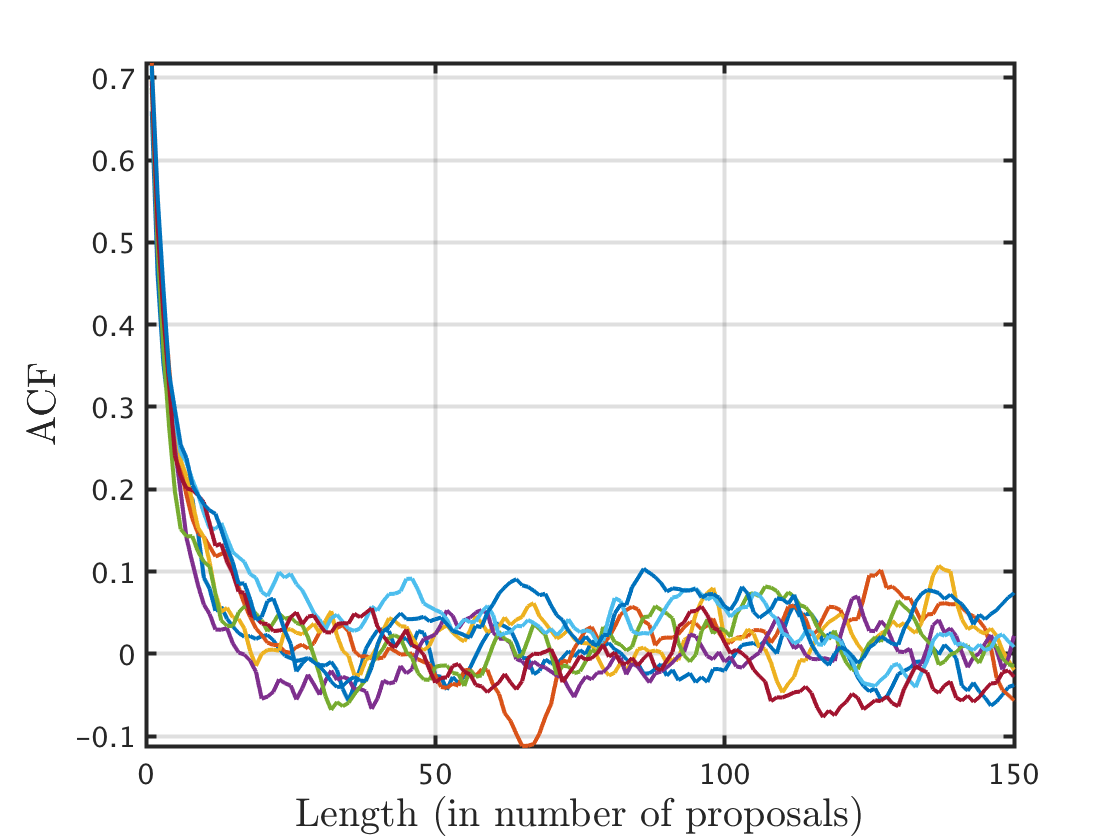}
  \caption{The ACF showing the longest autocorrelation length across all dimensions of MCMC is illustrated here. Monte Carlo samples of wrinkles are obtained by subsampling every $\Lambda = 100$ samples.}
  \label{fig:autocorrelation}
\end{figure}

\begin{figure}
  \centering
  \includegraphics[width=0.5\linewidth]{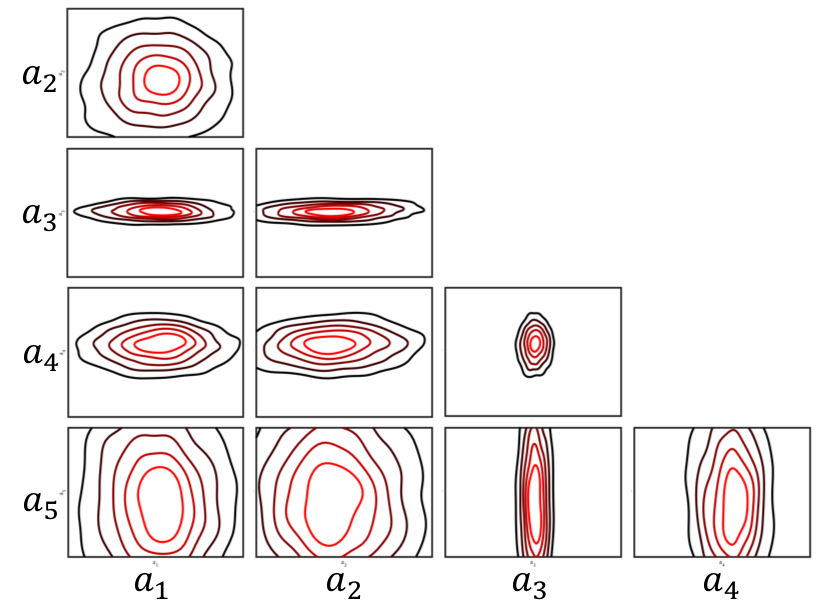} \includegraphics[width = 0.45\linewidth]{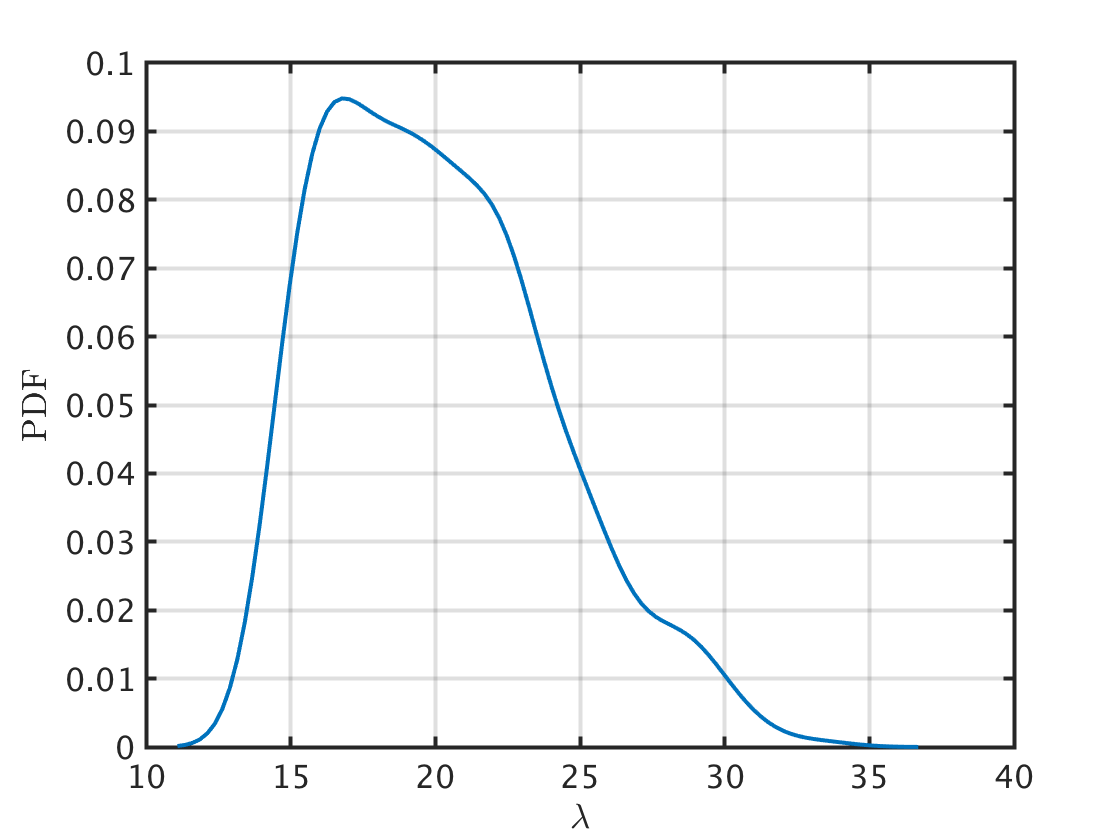}
  \caption{(Left) Two-dimensional posterior distributions of the first five coefficients $a_i$ in \cref{eqn:wrinkleDef}, note 2-D plot axes are plotted on a scale of $\pm 0.25$ to visualize dependencies. For example, the $a_3$ plots suggest that a relatively constant amount of the $3^{rd}$ KL mode compared to others is present in all wrinkles studied here. (Right) Posterior distribution of covariance length scale parameter $\lambda$ plotted separately.}
  \label{fig:plotmatrix}
\end{figure}

\begin{figure}
  \centering
  \includegraphics[width = \linewidth]{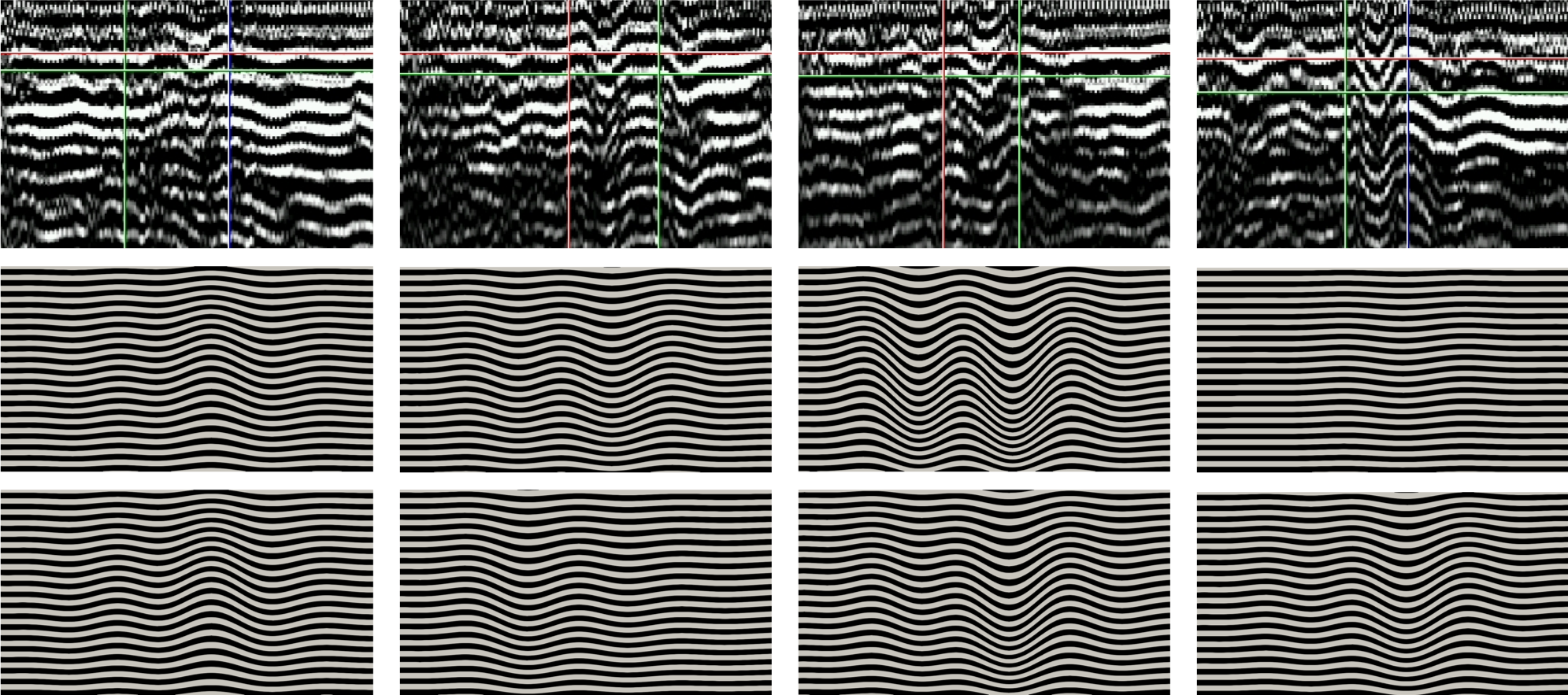}
  \caption{Top row shows B-Scan data, bottom two rows show $8$ independent posterior samples of wrinkles in B-Scan coordinates.}
  \label{fig:posteriorWrinkles}
\end{figure}

To improve the exploration of the posterior space we initialize five independent Markov chains. For the pCN proposal distribution \eqref{eq:proposal} we take $\beta = 0.25$ and $\sigma_{PCN} = 1$. These values were tuned to give an acceptance ratio of approximately $30\%$ as is widely suggested \cite{roberts1997}. We first estimate the integrated autocorrelation time $\Lambda$ for each chain. From \cref{fig:autocorrelation}, we note that  $\Lambda < 100$ in all cases. Random starting positions are sampled from the prior, then each chain is `burnt-in' over $10\Lambda \approx 1000$ MCMC steps. Post burn-in, each chain is subsampled at intervals of $2\Lambda$ till $N_{MC} = 200$ independent MCMC samples (or $8,000$ dependent samples per chain) are obtained. Posterior distributions of the first five coefficients $a_i$ from the combined dataset of all chains are visualized in \cref{fig:plotmatrix}. \cref{fig:posteriorWrinkles} shows a subset of 8 wrinkles out of the $200$ samples from the posterior distribution along with the four B-scans.

\subsection{Monte Carlo simulations}
\label{sec:fesim}

As a benchmark we first calculate the CBD for a pristine part, from which we calculate $M_c^\star = 8.93$ kNmm/mm. We also calculate knock downs for each of the B-Scan samples, by using the maximum \emph{a priori} (MAP) estimates for each scan
\begin{equation}
  M_d = \{8.61, 8.88, 8.62, 8.91 \} \; \mbox{kNmm/mm}
\end{equation}
Using the MCMC methodology we generate $N_{MC} = 200$ independent samples from the posterior distributions. All simulations were carried out on $400$ cores of the HPC cluster \texttt{Balena}, taking approximately 6 minutes per sample. The cluster comprises 192 nodes, each with two 8-core Intel Xeon E5-2650v2 Ivybridge process running at 2.6 GHz. Therefore total core time was approximately $20$ hours of computation. In practice, simulation time was less since a number of samples could be run in parallel by using the cluster's $\sim 3000$ available cores.

From these samples we estimate a mean of $\mathbb E[M_c] \approx 8.72$ kNmm/mm, equating to an average knock down of $2.4\%$. With a variance of $\mathbb V[M_c] \approx 0.094$ the $200$ samples we estimate the $95\%$ one-sided confidence interval of $0.053$ ($0.6\%$ of the mean value). Given that the finite element error at the mesh resolution chosen is approximately $0.5\%$ (as taken from \cite{dunecomposites}), the number of samples is sufficient to estimate the mean at the same accuracy as the discretization error given by the finite element model. Therefore no further samples were generated.

\begin{figure}
  \centering
  \includegraphics[width=0.49\linewidth]{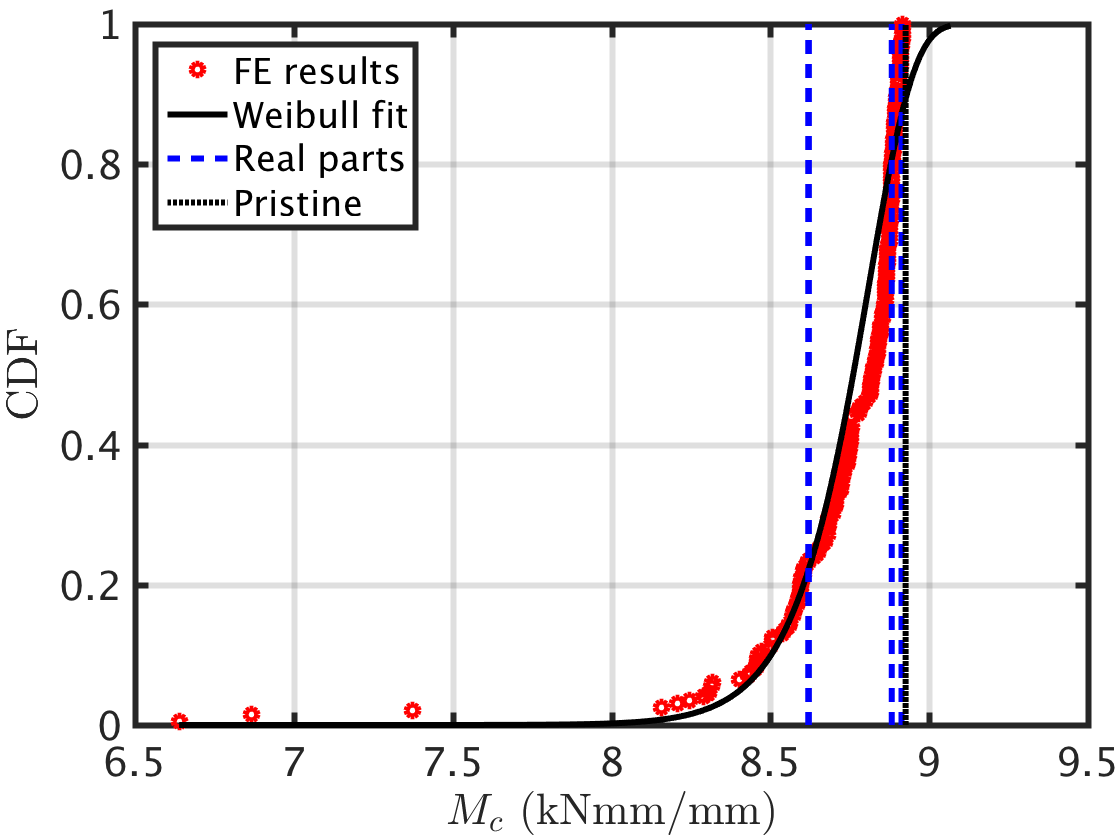}
  \includegraphics[width=0.49\linewidth]{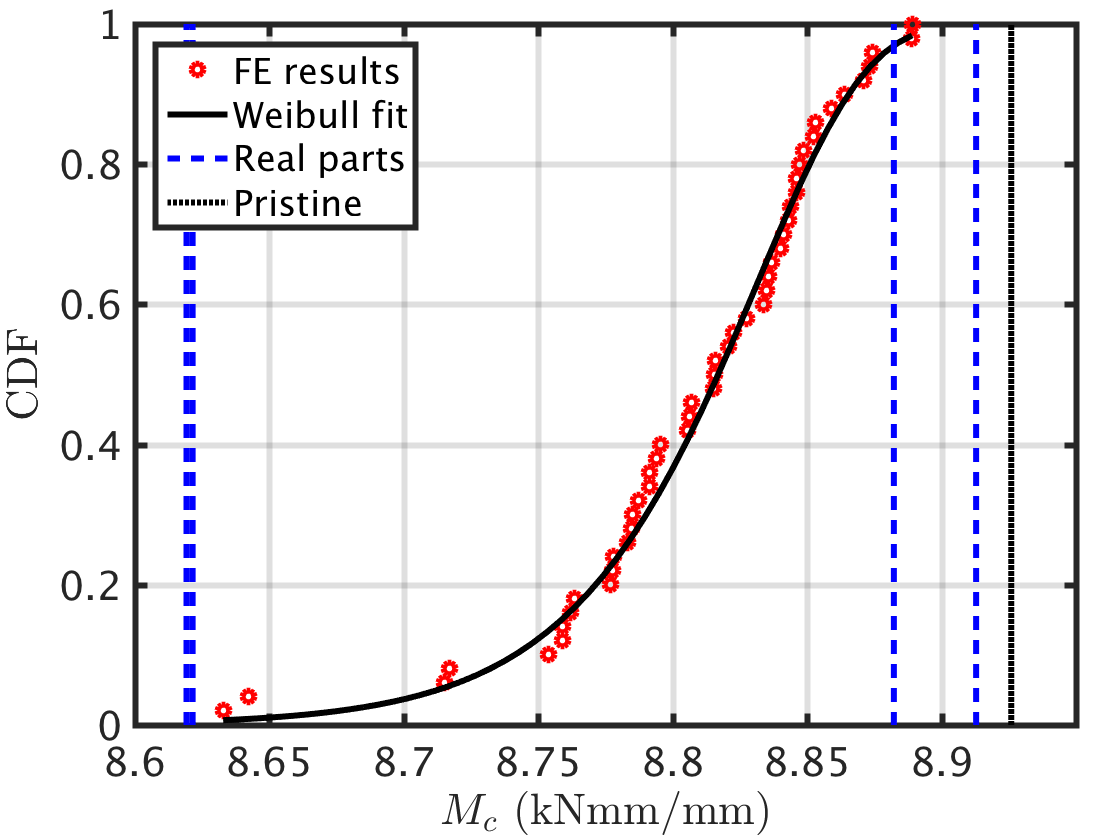}
  \caption{(Left) CDF of critical or failure moment $M_c$ per unit width of a part, where wrinkle distributions are using the Bayesian framework introduced within this paper. (Right) $M_c$ of samples obtained by assuming a Gaussian prior with mean and variance derived from data.}
  \label{fig:cdf}
\end{figure}

Whilst the mean seems like a small deviation from pristine strength, the worst of $200$ sample knocks the strength down by $26\%$. Therefore, rather than the mean itself we a more interested from an engineering viewpoint in the distribution of strength, particularly in the tails of the distribution, which represent larger knock downs in strength. \Cref{fig:cdf} (left) shows the CDF of strength distribution, along with Weibull fits; a common engineering way of quantifying material variability.

The Weibull model for failure assumes that fracture initiates at the weakest link. As \cref{fig:corr} shows the existence of a strong non-linear correlation between the maximum gradient of a wrinkle and its corresponding $M_c$, fitting a Weibull curve to the CDF is an appropriate choice. The Weibull curve, defined by \cref{eq:wbl}, predicts the probability of failure $P$;
\begin{equation}
  \label{eq:wbl}
  \mathbb P(M_c|M_W,M_S) = 1 - \exp \bigg[ - \bigg( -\frac{M_c}{M_S} \bigg)^{M_W} \bigg]
\end{equation}
where $M_c$ is the critical moment and $M_S$ is a scale parameter. Perhaps, the most important parameter is the Weibull modulus $M_W$ that can be thought of as a dispersion of defects in a part. A high dispersion is interpreted as an \emph{amorphous} presence of defects thus lacking a clear origin of failure. As a result, the loading bandwidth over which all parts fail is relatively narrow. Conversely, a low $M_W$ means that defects are concentrated in certain regions in such way that failure usually originates from these hot spots.

\Cref{fig:cdf} provides a comparison between the distributions obtained by sampling using the Markov chain methodology (left) in contrast to normal sampling (right) whereby coefficients $\boldsymbol{\xi}^{(k)}$ are drawn at random from a normal distribution centered about the observed mean with population standard deviation corrected for $n-1$ degrees of freedom with 95\% confidence bounds. For samples shown in \cref{fig:cdf} (left) the Weibull modulus $M_w = 62.9$. On the other hand, for samples shown in \cref{fig:cdf} (right), the Weibull modulus was found to be $M_w = 218.6$. The higher modulus indicates an even more uniform spread of softer regions. The modulus demonstrates the over conservative nature of normal sampling. It does not take into account the intrinsic correlation structure of the coefficients by assuming all coefficients are independent of each other. As a result, wrinkles generated this way are not likely to resemble observed defects, therefore, Markov chain sampling is preferred. This is an important result because it further strengthens the case for application of Bayesian methods to the wrinkle problem by clearly demonstrating the conservative nature of the current approach.

\subsection{An `engineering model' for Corner Bend Strength}

In this section we describe how the Bayesian methodology alongside a finite element model can be used to derive a distribution of corner bend strength due to random wrinkle defects, which in turn is parameterized by a Weibull model. We show how the results from these large finite element calculations can be distilled into a much simpler engineering approach or `look up' model, which provides some practical means of assessment on determining the influence of an observed wrinkle.

\begin{figure}[H]
  \centering
  \includegraphics[width=0.32\linewidth]{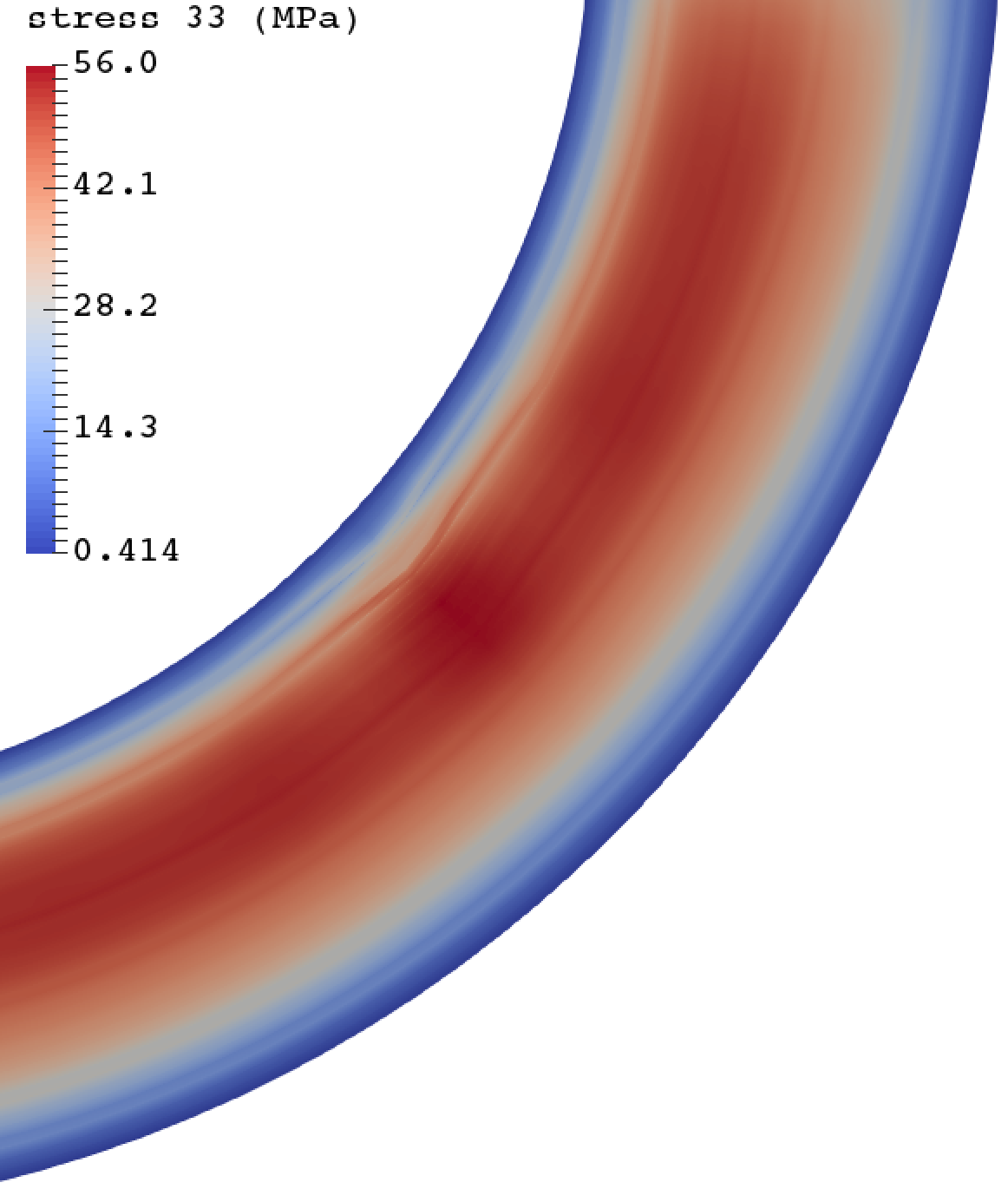}
  \includegraphics[width=0.32\linewidth]{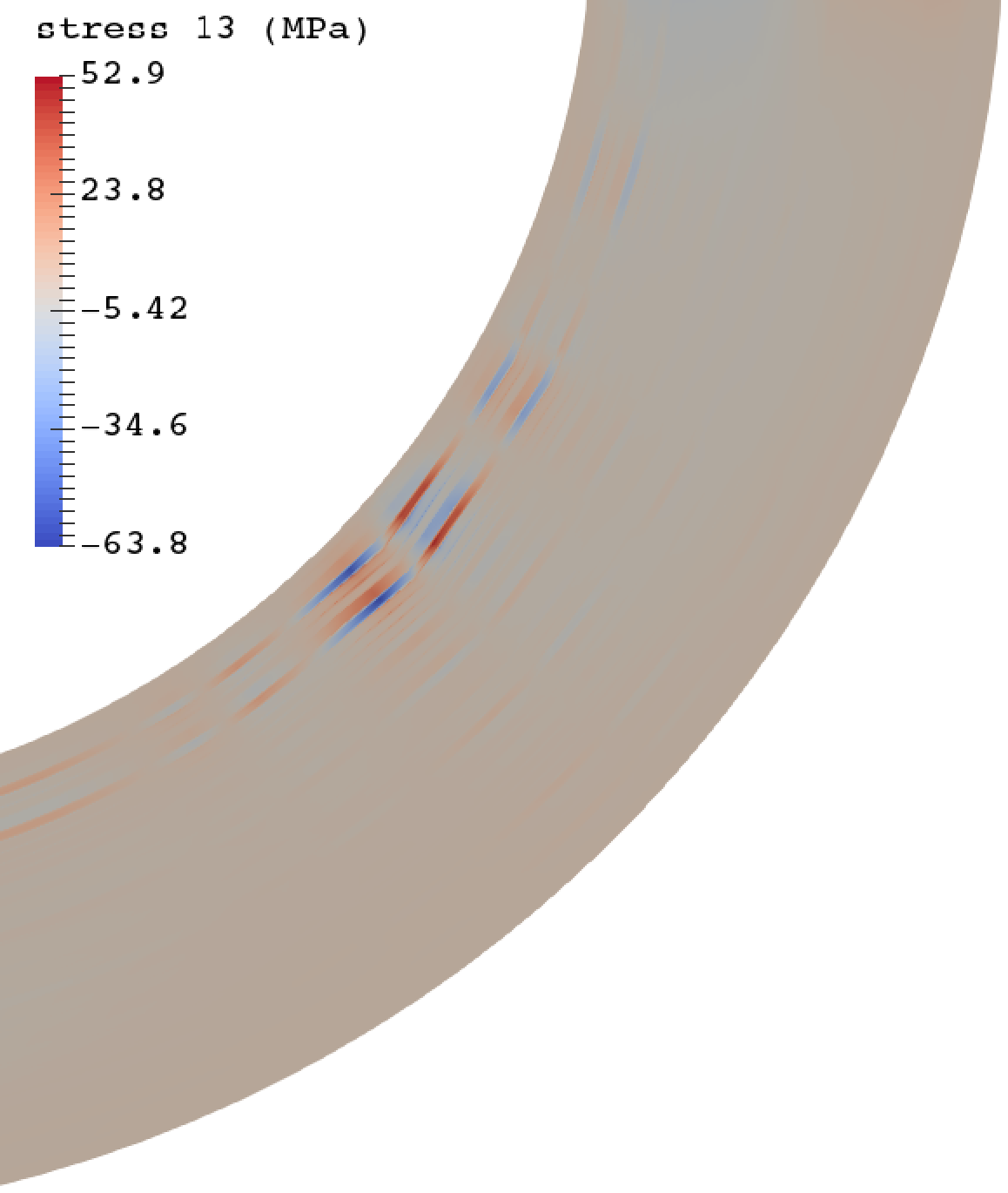}
  \includegraphics[width=0.32\linewidth]{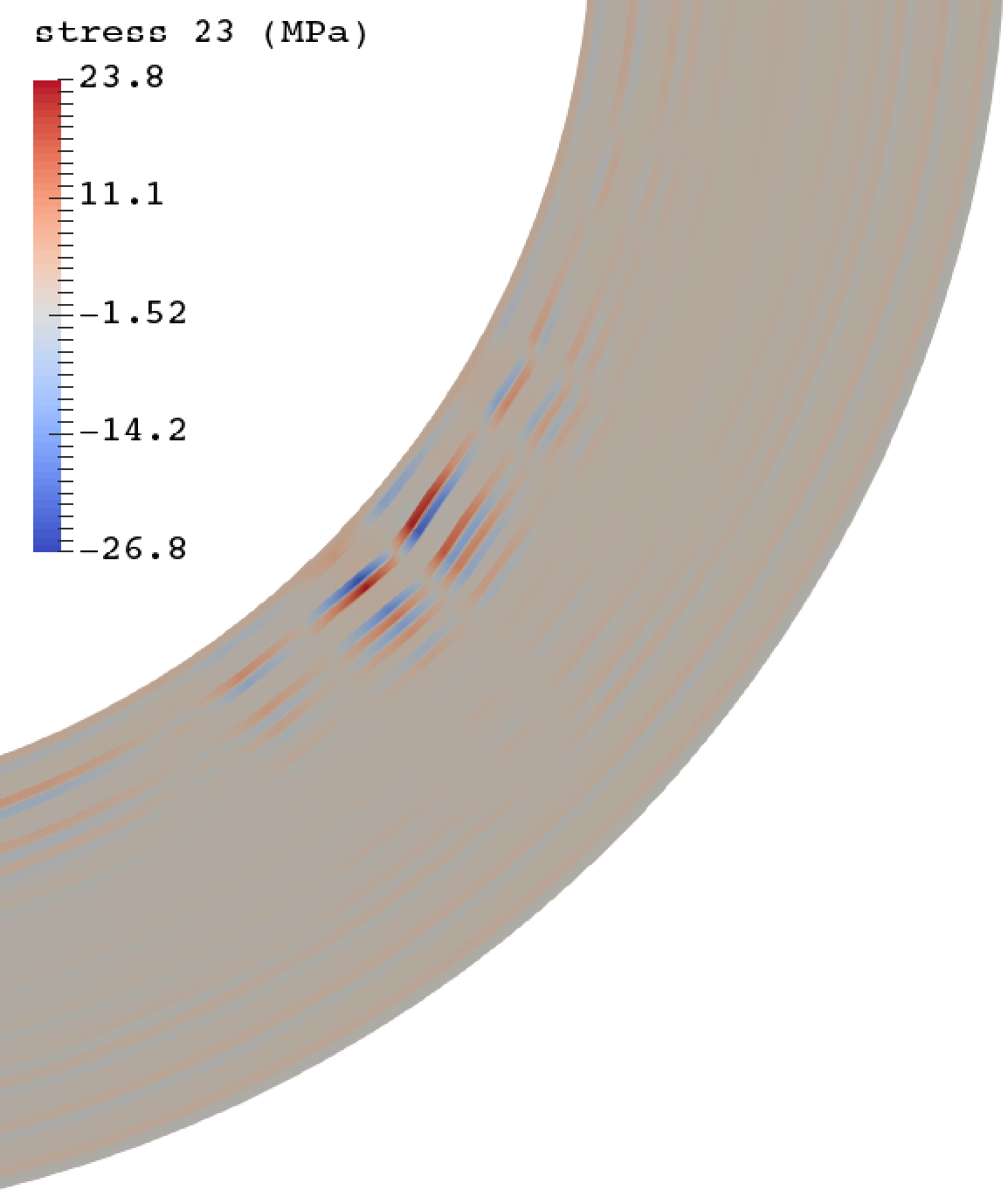}
  \caption{Left to right showing $\sigma_{33}$, $\tau_{13}$ and $\tau_{23}$ respectively at $M_a = 1$ kNmm/mm.}
  \label{fig:stressplots}
\end{figure}

\Cref{fig:stressplots} shows the wrinkle extracted from the B-scan in \cref{fig:cornerbend} embedded into the corner bend sample. The magnitude of the wrinkle in the B-scan is misleading since the pixels represent a length approximately 8 times larger in the vertical in comparison to the horizontal. In reality, the wrinkle is much smaller as shown in stress plots.

\begin{figure}
  \centering
  \includegraphics[width=0.55\linewidth]{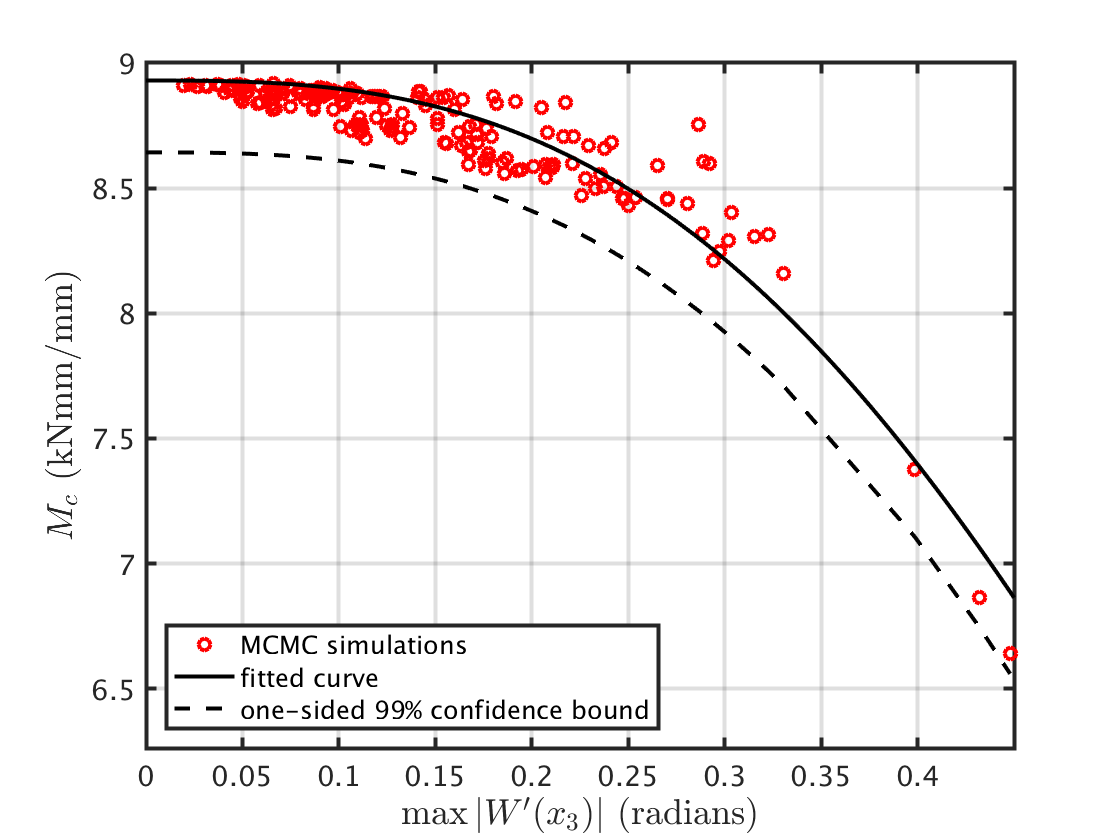}
  \caption{Approximating the relation between $M_c$ and maximum wrinkle slope with \cref{eq:prediction}. $q = 2.867$ and $\lambda_q= 4.212$ for the fitted curve. $q = 2.587$ and $\lambda_q = 3.834$ for lower $99\%$ confidence bound.}
  \label{fig:corr}
\end{figure}

An inspection of \cref{fig:stressplots} shows that interlaminar tensile stress ($\sigma_{33}$) concentrations form at peak wrinkle curvature whereas interlaminar shear ($\tau_{13}, \tau_{23}$) is concentrated around maximum wrinkle slope. Moreover, the allowable tensile stress ($s_{33}$), $60\%$ lower than allowable shear ($s_{13},s_{23}$), lends a greater contribution to $\mathcal{F}(\sigma)$ in \cref{eq:failure}. We therefore explore the correlation between maximum slope a knock down in strength. For this we parameterize the slope-failure dependency with an exponential relationship defined by
\begin{equation}
  \label{eq:prediction}
  M_c = M_c^\star \exp \bigg[ - \frac{W^\prime(x_3)^q}{\lambda_q} \bigg]
\end{equation}
where $M_c^\star$ is the strength of the pristine part. Model parameters, $q = 2.867$ and $\lambda_q= 4.212$ for the fitted curve in \cref{fig:corr}, enable prediction of critical moment for a given maximum gradient. The lower $99\%$ confidence bound for that prediction is computed with $q = 2.587$ and $\lambda_q = 3.834$.

\section{Conclusions and future work}
\label{sec:conclusions}
This paper proposes a generalized framework to quantify the effects of wrinkles in large composite structures. It combines ideas from NDT, image processing, Bayesian inference and FE modelling to create a rigorous methodology for visualizing, parameterizing and computing strength of wrinkles. The methodology is demonstrated through an industrially motivated case study with field data where we show the over conservative nature of the current design approach in comparison to the Bayesian, data-driven method to strengthen the certification by simulation idea.

Two dimensional ultrasonic scans (B-scans) are used to visualize wrinkles that form inside manufactured parts. They are parameterized using a Karhunen-Lo\'{e}ve basis due to their suitability for capturing multiple localized features. A possibly true distribution is inferred from observed wrinkles. The forward model - a finite element model - then determines the strength of a composite corner bend with a wrinkle embedded in it. Evaluating the forward model for a variety of defects, elucidates knockdown distribution. Whilst we focus on the influence of wrinkle defects, the general framework could be readily applied to other types defects, for example porosity provided an adequate basis is selected.

The theoretical strength for 200 independent MCMC wrinkles is evaluated to give an expected value approximately $2\%$ lower than the pristine strength. The same test for normally sampled wrinkles gives a much poorer estimate as the results fail to capture some of the observed wrinkles. The worst MCMC case however, suffers from a knockdown of approximately $26\%$. The cumulative strength distribution is well approximated by a Weibull curve with a relatively high Weibull modulus which is interpreted as the lack of a dominant failure mechanism or origin. In other words, it is difficult to find one particular wrinkle parameter universally responsible for failure.

An engineering model is constructed based on the significant negative correlation found between maximum gradient and critical moment of failure. Due to it's non-linearly decreasing nature, the gradient-failure relationship is parameterized by a negative exponential to produce a directly usable look up chart to estimate knockdown of a particular wrinkle.

We emphasize that this work demonstrates a method and the selected case study represents a narrow bandwidth of possible wrinkles due to a small set of observations. The available training data is a set of scans of pronounced wrinkles only, which makes it impossible to deduce their probability of occurrence in the first place. Consequently, all parts simulated here have wrinkles and suffer some strength knockdown. We do not claim our algorithm provides minimal error since the limited data set restricts the performance of the algorithm - a problem expected to be mitigated by a richer data set. Instead, we argue that industry design standards may now be challenged or reformulated. Moreover, there remains some room to develop application specific bases and more accurate estimates of decay functions and location parameters. In this way, building true representations of the parameters at the coupon level affords us a sampling space from which defects can be generated and embedded into much larger components models.

\section{Bibliography}



\bibliographystyle{model1-num-names}
\bibliography{references}







\end{document}